\documentclass[final,3pt,scale=0.8,times,12pt]{elsarticle}
\usepackage{CJK} 
\biboptions{numbers,sort&compress}
\usepackage{graphicx,enumitem} 
\usepackage[margin=0.75in]{geometry}
\usepackage{amssymb}
\usepackage{amsmath}
\usepackage{subfigure}
\usepackage{epstopdf}
\usepackage{bibentry}
\usepackage{caption}
\usepackage{color}
\usepackage{mathrsfs}
\usepackage{multirow}
\usepackage{makecell}
\usepackage{hhline}
\usepackage{breqn}
\usepackage{lineno,hyperref}
\usepackage{fancyhdr}
\usepackage{geometry}
\usepackage{hyperref}            
\hypersetup{
    colorlinks=true,
    linkcolor=RoyalBlue,
    citecolor=ForestGreen,
    urlcolor=Orchid
}
\journal{***}
\DeclareMathOperator{\sech}{sech}
\numberwithin{equation}{section}

\newtheorem{prop}{Proposition}[section]

\newtheorem{rk}{Remark}[section]
\allowdisplaybreaks[4]

\def\e{{\rm e}}

\def\e{{\rm e}}

\begin{document}
\begin{frontmatter}

\title{\textbf{Localized stem structures in soliton reconnection of the asymmetric Nizhnik-Novikov-Veselov system}}

\author[a]{Feng Yuan\corref{cor2}}
\author[b]{Jingsong He\corref{cor1}}
\author[b]{Yi Cheng}
\address[a]{College of Science, Nanjing University of Posts and Telecommunications, Nanjing, 210023, P. R. China}
\address[b]{Institute for Advanced Study, Shenzhen University, Shenzhen, 518060, P. R. China}
\address[b]{School of Mathematical Sciences, USTC, Hefei, Anhui 230026, P. R. China}

\cortext[cor1]{Corresponding author. E-mails: hejingsong@szu.edu.cn}
\cortext[cor2]{Corresponding author. E-mails: fengyuan@njupt.edu.cn}

\begin{abstract}
The reconnection processes of 3-solitons with 2-resonance can produce distinct local structures that initially connect two pairs of V-shaped branches, then disappear, and later re-emerge as new forms. We call such local structures as stem structures. In this paper, we investigate the variable-length stem structures during the soliton reconnection of the asymmetric Nizhnik-Novikov-Veselov system. We consider two scenarios: weak 2-resonances (i.e., $a_{12}=a_{13}=0,\,0<a_{23}<+\infty$) and strong 2-resonances (i.e., $a_{12}=a_{13}=+\infty,\,0<a_{23}<+\infty$). We determine the asymptotic forms of the four arms and their corresponding stem structures using two-variable asymptotic analysis method which is involved simultaneously with one space variable $y$ (or $x$) and
one temporal variable $t$. Different from known studies, our findings reveal that the asymptotic forms of the arms $S_2$ and $S_3$ differ by a phase shift as $t\to\pm\infty$. Building on these asymptotic forms, we perform a detailed analysis of the trajectories, amplitudes, and velocities of the soliton arms and stem structures. Subsequently, we discuss the localization of the stem structures, focusing on their endpoints, lengths, and extreme points in both weak and strong 2-resonance scenarios.
\end{abstract}


\end{frontmatter}

\section{Introduction}
Soliton theory is an important research area in the nonlinear science, focusing on a particular type of nonlinear wave known as a solitary wave. The solitary wave was first observed by J. S. Russell \cite{russel} in a canal and later explained by the KdV equation formulated by Korteweg and de Vries \cite{kdv1895}. After colliding, two solitary waves can return to their original shapes and conserve their momentums and energies, which satisfies the properties of elastic collisions \cite{prl1965,cpam1974,book03,book04,interaction03}. This ``particle-like" behavior of solitary waves is a crucial characteristic corresponding to their naming term ``soliton".

In physics, solitons emerge as distinctive manifestations of nonlinear phenomena. Endowed with unique attributes including conservation of energy, momentum preservation, and stability, solitons find extensive application across diverse fields of physics \cite{book04}. Mathematically, solitons are defined as stable and energy-conserving non-dispersive solutions within specific nonlinear partial differential equations. Soliton equations serve as mathematical models for describing the essences and characteristics of solitons. In a seminal contribution in 1965, Zabusky and Kruskal introduced the concept of solitons for the KdV equation through numerical method \cite{prl1965}. Two years later, a method of solving the initial value problem of the KdV equation was proposed in Ref.\ \cite{prl1967}, which also (or further) produces a closed form of $n$ soliton solutions. This method is now widely known as inverse scattering method. Subsequently, in 1974, the exact and analytical $n$-soliton solutions for the KdV equation were successfully derived utilizing inverse scattering method \cite{cpam1974}. Since then, numerous soliton equations have been discovered \cite{book03,book04,prl1965}, and other analytical methods have been developed for solving these equations, including the Darboux transformation \cite{dt01,dt02,dt03}, and the Hirota bilinear method \cite{hirota1971,book01}.

The study of soliton interactions is a key research focus in soliton theory. While elastic collisions are a fundamental feature of solitons, solitons may also undergo inelastic collisions \cite{interaction04,interaction05,interaction06} under special conditions. A distinctive class of ``resonant" interactions arises when the wave numbers and frequencies of solitons meet specific constraints, first identified in the context of the Kadomtsev-Petviashvili (KP) equation \cite{stem01,interaction16,interaction17,interaction07,jpsj1983-1}. Resonance is generated through two main mechanisms: (i) resonance occurs when the intersection angle of solitons meets certain conditions \cite{jpsj1980}; (ii) resonance arises when the phase shift becomes indefinite \cite{rao03,interaction09,interaction12}. In this study, we examine soliton interactions from the perspective of phase shift: elastic collisions exhibit a finite phase shift, while resonant collisions are characterized by an infinite phase shift. Furthermore, when the phase shift in an elastic collision is finite but approaches infinity, it is termed as a quasi-resonant collision.

Commonly speaking, two-dimensional line solitons extend infinitely in space. However, quasi-resonant and resonant collisions can give rise to localized stem structures. The concept of the stem structure was first introduced in the context of Mach reflection \cite{stem01,stem02}. In this phenomenon, the apex of the incident and reflected wave separates from the wall and is connected to it by a third solitary wave that perpendicularly intersects the wall, now widely known as the stem wave. In the quasi-resonant state, the vertices of the X-shaped solitons are significantly separated due to the phase delay, forming two V-shaped solitons connected by a new localized structure, as another instance of the stem structure \cite{stem1980}. Additionally, the 3-soliton can produce a localized structure when it resonates twice. In this scenario, the soliton appears as four surrounding infinity arms connected by an intermediate soliton (stem structure). Over time, the middle soliton gradually shrinks, resulting in the merge of the endpoints of the other four arms. They later separate again and are reconnected by a new middle soliton with a different orientation. The entire process is known as the soliton reconnection \cite{interaction08,interaction25,interaction12}. The common features of both quasi-resonant and soliton reconnection are the localization and finite length of the stem structures.

In this study, we define the stem structure (or wave) as a localized wave connecting the vertices of two pairs of V-shaped line solitons. Spatial localization is the fundamental characteristic of the stem structure. Investigating the stem structure in solitons can provide deeper insights into the nature of solitons and enhances our understanding of various nonlinear phenomena. The stem structure has been studied to some extent in solitons by a graphic way without analytic formulas of the stem structures. For example:

\begin{itemize}
\setlength{\itemsep}{-5pt}
  \item In the case of 2-solitons for the KP equation and the Boussinesq equation, the central region of the X-shaped soliton forms a constant-length stem when the phase shift is finitely large \cite{jpsj1980,interaction09,interaction26}.
  \item For 3-solitons in the KP and Davey-Stewartson (DS) equations, soliton reconnection generates two finite structures: during this process, one finite structure gradually disappears while another emerges \cite{interaction08,interaction12}.
\end{itemize}

Although various equations describe the two types of soliton collisions resulting from interactions \cite{interaction09,interaction10,interaction11,interaction13,interaction14}, there has been limited analysis of the stem structure of solitons, aside from some intuitive and interesting graphical demonstrations. In this work, we choose the asymmetric Nizhnik-Novikov-Veselov (ANNV) system to analytically study the localized stem structures in 3-solitons, primarily due to its significant applications in numerous fields of physics, especially in shallow waves driven by weakly nonlinear restoring forces in incompressible fluids \cite{annv04,annv02}. This system was originally introduced by Boiti utilizing the weak Lax pair \cite{annv01}, and it has the following form:
\begin{equation}\label{annveq}
	\left\{
	\begin{aligned}
		&u_t+v_{xxx}=3(uv)_x,\\
		&u_x=v_y.\\
	\end{aligned}
	\right.
\end{equation}
Here, $u(x,\,y,\,t)$ and $v(x,\,y,\,t)$ denote the dimensionless velocity components. 
The spectral transformation for this system has been investigated in Ref.\ \cite{annv01,book04}. 
It has been shown that when $x=y$, Eq.\ \eqref{annveq} simplifies to the KdV equation \cite{annv01}. 
Additionally, this system can be derived through the inner parameter-dependent symmetry constraint of the KP equation \cite{annv04}. 
Clarkson and Mansfield have investigated several crucial aspects of the ANNV system, including the Painlev\'{e} property and similarity solutions \cite{annv05}. 
Furthermore, a plethora of solution types has been delineated in the past researches. Dromion and kink solutions have been meticulously crafted and documented in the seminal works of \cite{annv02,annv16}. 
Notably, the quasi-periodic solutions have been rigorously formulated in the scholarly discourse outlined in Ref.\ \cite{annv07}. 
Similarly, soliton solutions have been meticulously derived and discussed in the scholarly investigation chronicled in Ref.\ \cite{annv10}. 
Moreover, lump solution has been exhaustively examined and elucidated in the comprehensive research efforts documented in Refs.\ \cite{annv12}. 
Finally, the elucidation of rational and semi-rational solutions has been systematically undertaken in the insightful analysis presented in Refs.\ \cite{annv09,annv15}.

Recently, the quasi-resonant 2-solitons with constant-length stem structures have been constructed and studied in Ref.\ \cite{yuan2024}. 
However, the 2-resonant 3-soliton solution of Eq.\ \ref{annveq}, which produces a variable-length stem structure during soliton reconnection, has not been explored. 
The essential challenges in this study on stem structures are as follows:
\begin{itemize}
\setlength{\itemsep}{-5pt}
\item How to consider the asymptotic analysis  involved simultaneously with one space variable $y$ (or $x$) and
one temporal variable $t$ in two dimensional solitons?  
There are many known results on the asymptotic analysis just on only one spatial variable ($x$ or $y$) \cite{jpsj1983-1,pla2023} or on time variable $t$ \cite{prsa2021,rao01}, which cannot provide the accurate form of  soliton arms. 
This two-variable asymptotic analysis method is crucial to determine ends of stem structures and find the accurate form of soliton arms before and after the interactions of multi-soliton.
\item Based on the accurate form of soliton arms, how to find the analytical forms for the ends, length, trajectory of the stem structures?
\end{itemize}

The primary objective is to overcome the above challenges and to study the variable-length stem structure in 2-resonant 3-soliton solutions. 
Unlike the resonant solution derived in Ref.\ \cite{interaction12} by transforming the polar coordinates of $x$ and $y$ and subsequently controlling the angle between them, our study induces soliton reconnection by satisfying the 2-resonance condition for two of the three $a_{ij}$. 
The solutions obtained in our paper resemble the partially resonant solutions described in Ref.\ \cite{interaction12} (subsection 7.2). 
We will demonstrate that the 2-resonance condition \cite{jpsj1983-1} implies that resonance will occur at two different points on the $(x, y)$-plane. 
Furthermore, soliton reconnections manifest in two cases: weak and strong resonances, distinguished by whether $a_{ij}=0$ or $a_{ij}=\infty$. 
It is worth noting that due to the extremely complex interaction mechanism of multi-solitons near $t=0$, which has not yet been fully characterized, 
the asymptotic analysis of 2-resonant 3-solitons and the research on stem structures conducted in this paper are only applicable to the cases of $t \ll 0$ and $t\gg 0$.
The organization of the paper is as follows: In Sec. \ref{sec2}, we recall the expressions and N-solution for the ANNV equation \eqref{annveq} based on the Hirota Bilinear method. 
In Secs. \ref{sec3} and \ref{sec4}, we delve into the asymptotic forms and the stem structure in 3-soliton solutions generated by weak 2-resonance and strong 2-resonance, respectively, exploring soliton reconnection and studying the properties of the variable-length stem structures produced during reconnection. 
Finally, in Sec. \ref{summary}, we provide conclusions and discussions for this paper.

\section{Recall the N-soliton solutions of the ANNV system}\label{sec2}
The N-soliton solutions of \eqref{annveq} generated by the Hirota Bilinear method have been given by \cite{annv06}
\begin{align}
&u=-2(\ln f)_{xx},\,v=-2(\ln f)_{xy},\label{uv}\\
&f^{[N]}=\sum_{\mu =0, 1} \exp \left(\sum_{i<j}^{N} \mu_{i}\mu_{j}A_{ij}+ \sum_{i=1}^{N} \mu_{i}\xi_{i}\right),\label{nf}
\end{align}
where,
\begin{flalign}\label{etakpw}
	\begin{split}
		\xi_j=k_jx+p_jy-k_j^3t+\xi_j^0,\,\exp(A_{ij})=\frac{(k_i-k_j)(p_i-p_j)}{(k_i+k_j)(p_i+p_j)}\triangleq a_{ij}\geqslant 0.
	\end{split}
\end{flalign}

By substituting $N=3$ into Eq. \eqref{nf}, the order-3 soliton solution can be derived using \eqref{uv} and the subsequent equation:
\begin{flalign}\label{3soliton}
	\begin{split}
		f^{[3]}&=1+\exp\xi_1+\exp\xi_2+\exp\xi_3+a_{12}\exp(\xi_1+\xi_2)+a_{13}\exp(\xi_1+\xi_3)\\
&+a_{23}\exp(\xi_2+\xi_3)+a_{12}a_{13}a_{23}\exp(\xi_1+\xi_2+\xi_3).
	\end{split}
\end{flalign}
The phase shift of the 3-soliton solution is denoted as $\Delta_{ij}=\ln a_{ij}$ for $(i,\,j=1,\,2,\,3 \text{ and } i<j)$. 
Different conditions on the phase shift give rise to distinct types of collisions between the three solitons: 
elastic collisions for $\Delta_{ij}<\infty$, and resonant collisions for $\Delta_{ij}=\infty$. 
In this paper, we narrow our focus to the variable-length stem structure in 3-solitons by 2-resonance, occurring when $a_{12}=a_{13}=0$ or $a_{12}, a_{13}=\infty$. Here, the 2-resonance condition denotes that resonance can occur twice at different points on the $(x,\,y)$-plane. The weak resonance corresponds to $a_{ij}=0$, and the strong resonance corresponds to $a_{ij}=+\infty$.

\begin{rk}
The distinction between strong and weak resonances lies in their outcomes: 
strong resonance between \(S_i\) ($f_i=1+\exp\xi_i$) and \(S_j\) ($f_j=1+\exp\xi_j$) produce a soliton \(S_{i+j}\) ($f_{i+j}=1+\exp(\xi_i+\xi_j)$), whereas weak resonance yield \(S_{i-j}\)  ($f_{i+j}=1+\exp(\xi_i-\xi_j)$).
\end{rk}

\begin{rk}
The interaction mechanism of multi-solitons in the neighborhood of $t=0$ is extremely complex and has not yet been fully described. 
Based on this, the research on the asymptotic forms and stem structures of 2-resonance 3-solitons carried out in this paper only holds under the conditions of $t\ll 0$ and $t\gg 0$.
\end{rk}

\begin{table}
  \centering
  \caption{Physical quantities of the soliton}
  \label{tab:t1}
  \begin{tabular}{cccccc}
    \Xhline{1pt}
     Soliton & Trajectory & Velocity & Amplitude & Components \\
    \hline
    \multirow{2}{*}{$S_{j}$} & \multirow{2}{*}{$l_j$} & \multirow{2}{*}{$(k_j^2, -\frac{k_j}{p_j})$} & $-\frac{k_jp_j}{2}$ & $u_{j}$ \\
     & & &$-\frac{k_j^2}{2}$ & $v_{j}$ \\
    \hline
    \multirow{2}{*}{$S_{j}$} & \multirow{2}{*}{$\widehat{l_j}$} & \multirow{2}{*}{$(k_j^2, -\frac{k_j}{p_j})$} & $-\frac{k_jp_j}{2}$ & $\widehat{u_{j}}$ \\
     & & &$-\frac{k_j^2}{2}$ & $\widehat{v_{j}}$ \\
    \Xhline{1pt}
  \end{tabular}
  \caption*{\captionsetup{justification=raggedright,singlelinecheck=false,format=hang} \quad The $j$-th soliton $S_j$ $(j=1,2,3)$ is composed by two components $u_j$ and $v_j$, and their essential properties are summarized in Table \ref{tab:t1}. The trajectories of the three kinds of solitons $S_j$ are given by \eqref{l01}.}
\end{table}
\section{The stem structure in 3-soliton generated by weak 2-resonance in soliton reconnection}\label{sec3}
In this section, we focus on the 3-soliton generated by weak 2-resonance in the case of $a_{12}=a_{13}=0$ and $0<a_{23}<+\infty$, during the soliton reconnection. The corresponding transformation of Eq.\ \eqref{3soliton} is captured by the expression:
\begin{flalign}\label{3soliton1}
\begin{split}
f_{weak}=1+\exp\xi_1+\exp\xi_2+\exp\xi_3+a_{23}\exp(\xi_2+\xi_3).
\end{split}
\end{flalign}
To delve into the 3-soliton solution with weak 2-resonance, as defined by Eqs.\ \eqref{3soliton1} and \eqref{uv}, we explore its asymptotic behavior by two-variable asymptotic analysis, which is involved with $y$ and $t$. In order to satisfy the condition $a_{12}=a_{13}=0,\,0<a_{23}<+\infty$, without loss of generality, we will discuss it in two cases: (1) $0<k_1=k_3<k_2,\,p_1=p_2>p_3>0$ and (2) $k_1=k_2>k_3>0,\,0<p_1=p_3<p_2$.

\subsection{The asymptotic analysis with $0<k_1=k_3<k_2,\,p_1=p_2>p_3>0$}
Previous researches on asymptotic analysis considering the asymptotic behaviors of the spatial variables and treats $t$ as a constant \cite{jpsj1983-1, pla2023,rao01,prsa2021}, producing only static asymptotic forms. Note that the parameters $c$ and $c_j$ introduced in the asymptotic analysis in this paper do not refer to specific values but rather to general constants.

First, we use $t$ as a constant as the asymptotic analysis in Refs.\ \cite{jpsj1983-1, pla2023,rao01,prsa2021}, in order to show clearly that the description of soltion arms is not accurate due to the different appearance of phase shift terms, by the known way. We consider the region $\zeta_j = x + \frac{p_j}{k_j}y= \text{constant}$:

(1) When $y \to -\infty$, we have $\eta_1 \approx c$, $\eta_2 \to +\infty$, $\eta_3 \to +\infty$, $\eta_2 \gg \eta_3$, and then
$f \approx \exp(\eta_2) + a_{23} \exp(\eta_2 + \eta_3)$.

(2) When $y \to +\infty$, we have $\eta_1 \approx c$, $\eta_2 \to -\infty$, $\eta_3 \to -\infty$, $\eta_2 + \eta_3 \to -\infty$, and then
$f \approx 1 + \exp(\eta_1)$.

(3) When $y \to -\infty$ we have $\eta_2 \approx c$, $\eta_1 \to -\infty$, $\eta_3 \to -\infty$, $\eta_2 + \eta_3 \to -\infty$, and then
$f \approx 1 + \exp(\eta_2)$.

(4) When $y \to +\infty$, we have $\eta_2 \approx c$, $\eta_1 \to +\infty$, $\eta_3 \to +\infty$, $\eta_2 + \eta_3 \to +\infty$, $\eta_1 \gg \eta_3$, and then
$f \approx \exp(\eta_1) + a_{23} \exp(\eta_2 + \eta_3)$.

(5) When $y \to -\infty$, we have $\eta_3 \approx c$, $\eta_1 \to -\infty$, $\eta_2 \to +\infty$, $\eta_2 + \eta_3 \to +\infty$, and then
$f \approx \exp(\eta_2) + a_{23} \exp(\eta_2 + \eta_3)$.

(6) When $y \to +\infty$, we have $\eta_3 \approx c$, $\eta_1 \to +\infty$, $\eta_2 \to -\infty$, $\eta_2 + \eta_3 \to -\infty$, and then
$f \approx \exp(\eta_1)$.

After sorting above analysis out, we have the following asymptotic forms,
\begin{flalign}
\begin{split}\label{asy00}
&f_2\sim 1+\e^{\eta_2},\,f_3\sim 1+a_{23}\e^{\eta_3},\,(y\to -\infty),\\
&f_1\sim 1+\e^{\eta_1},\,f_{1-2-3}\sim 1+a_{23}\e^{\eta_2+\eta_3-\eta_1},\,(y\to +\infty).
\end{split}
\end{flalign}

In fact, the asymptotic forms obtained here are not accurate and cannot reflect the temporal attributes of the asymptotic forms at $t\rightarrow \pm \infty$. The accurate asymptotic forms will be given in \eqref{asyf01} as $t\rightarrow -\infty$ and \eqref{asyf02} as $t\rightarrow +\infty$, which includes different appearance of phase shift term $a_{23}$. This difference of $u$ between $t$ goes to $\pm\infty$ is very crucial to understand the
asymptotic behavior of $u$. This verifies that asymptotic from in \eqref{asy00}, does not provide a accurate form of soliton arms, and shows it is necessary to use two-variable asymptotic method in this paper. Specifically, in a 2-resonant 3-soliton solution, some arms experience a phase shift as $t\to -\infty$ or $t\to +\infty$ due to the influence of $a_{23}$. In this paper, the asymptotic analysis we undertake necessitates a concurrent examination of both a spatial variable $y$ and a temporal variable $t$, a methodology referred to in the paper as two-variable asymptotic analysis method. To be convenient, we consider the region $\eta_j= x + \frac{p_j}{k_j}y - k_j^2t = \text{constant}$ instead of $\zeta_j = x + \frac{p_j}{k_j}y = \text{constant}$. The former allows $\eta_j$ to be treated as a constant for any variable, while the latter requires $t$ to be constant to treat $\zeta_j$ as a constant.

{\bf I.} On the region $\eta_1=x+\frac{p_1}{k_1}y-k_1^2t$, we have
$$\xi_1=k_1\eta_1+c_1,\,\xi_2=k_2\eta_1+\frac{k_1p_2-k_2p_1}{k_1}y+k_2(k_1^2-k_2^2)t+c_2,\,\xi_3=k_3\eta_1+\frac{k_1p_3-k_3p_1}{k_1}y+c_3,$$
$$\xi_2+\xi_3=(k_2+k_3)\eta_1+\frac{(p_2 + p_3)k_1 - p_1(k_2 + k_3)}{k_1}y+((k_2+k_3)k_1^2-(k_2^3+k_3^3))t+c_4.$$

{\bf (a)} In the case of $t\to +\infty$:  We have $\xi_1\approx c,\,\xi_2\to -\infty,\,\xi_3\approx c,\,\xi_2+\xi_3\to -\infty$, and then Eq.\ \eqref{3soliton1} becomes $f\sim 1+\e^{\xi_1}+\e^{\xi_3}$. Further we can get:
$$f\sim 1+\e^{\xi_1},\,y\to+\infty;\,f\sim 1+\e^{\xi_3},\,y\to-\infty;\,f\sim \e^{\xi_1}+\e^{\xi_3},\,x\to +\infty.$$

{\bf (b)} In the case of $t\to -\infty,\,y\to -\infty$:  We have $\xi_1\approx c,\,\xi_2\to +\infty,\,\xi_3\to +\infty,\,\xi_2+\xi_3\to +\infty$. Because of $\xi_2-\xi_3=(k_2-k_3)\eta_1+\frac{k_1p_2-k_2p_1-k_1p_3+k_3p_1}{k_1}y+k_2(k_1^2-k_2^2)t\to +\infty$ for $t\to -\infty,\,y\to -\infty$, we have $\e^{\xi_2}+a_{23}\e^{\xi_2+\xi_3}\gg\e^{\xi_3}$. And then we can get:
$$f\sim \e^{\xi_2}+a_{23}\e^{\xi_2+\xi_3},\,t\to -\infty,\,y\to -\infty.$$

{\bf (c)} In the case of $t\to -\infty,\,y\to +\infty$: By the limit $y\to +\infty$, we have $\xi_1\approx c,\,\xi_2\to -\infty,\,\xi_3\to -\infty,\,\xi_2+\xi_3\to -\infty$. Since $a_{23}$ is not applied to soliton $S_1$, we get $f\sim 1+\e^{\xi_1},\,y\to+\infty$ for even $t$.

{\bf II.}  On the region $\eta_2=x+\frac{p_2}{k_2}y-k_2^2t$, we have
$$\xi_1=k_1\eta_2+\frac{k_2p_1-k_1p_2}{k_2}y+k_1(k_2^2-k_1^2)t+c_1,\,\xi_2=k_2\eta_2+c_2,\,\xi_3=k_3\eta_2+\frac{k_2p_3-k_3p_2}{k_2}y+k_3(k_2^2-k_3^2)t+c_3,$$
$$\xi_2+\xi_3=(k_2+k_3)\eta_2+\frac{k_2p_3 - k_3p_2}{k_2}y+k_3(k_2^2-k_3^2)t+c_4.$$

{\bf (a)} In the case of $t\to +\infty$:  We have $\xi_2\approx c,\,\xi_1\to +\infty,\,\xi_3\to +\infty,\,\xi_2+\xi_3\to +\infty$, and then Eq.\ \eqref{3soliton1} becomes $f\sim \e^{\xi_1}+\e^{\xi_3}+a_{23}\e^{\xi_2+\xi_3}$. Further we can get:
$$f\sim \e^{\xi_3}+a_{23}\e^{\xi_2+\xi_3},\,y\to -\infty;\,f\sim \e^{\xi_1}+a_{23}\e^{\xi_2+\xi_3},\,y\to+\infty;\,f\sim \e^{\xi_1}+\e^{\xi_3},\,x\to-\infty.$$

{\bf (b)} In the case of $t\to -\infty,\,y\to -\infty$:  We have $\xi_2\approx c,\,\xi_1\to -\infty,\,\xi_3\to -\infty,\,\xi_2+\xi_3\to -\infty$, and then Eq.\ \eqref{3soliton1} becomes
 $$f\sim 1+\e^{\xi_2},\,t\to -\infty,\,y\to -\infty.$$

{\bf (c)} In the case of $t\to -\infty,\,y\to +\infty$: By the limit $y\to +\infty$, we have $\xi_2\approx c,\,\xi_1\to +\infty,\,\xi_3\to +\infty,\,\xi_2+\xi_3\to +\infty$ and $f\sim \e^{\xi_1}+\e^{\xi_3}+a_{23}\e^{\xi_2+\xi_3}$. Because of $\xi_3-\xi_1=(k_3-k_1)\eta_2+(p_3-p_1)y\to-\infty$, we have $\e^{\xi_1}+a_{23}\e^{\xi_2+\xi_3}\gg \e^{\xi_3}$. So $f\sim \e^{\xi_1}+a_{23}\e^{\xi_2+\xi_3},\,y\to +\infty$ for even $t$.

{\bf III.}  On the region $\eta_3=x+\frac{p_3}{k_3}y-k_3^2t$, we have
$$\xi_1=k_1\eta_3+\frac{k_3p_1-k_1p_3}{k_3}y+c_1,\,\xi_2=k_2\eta_3+\frac{k_3p_2-k_2p_3}{k_3}y+k_2(k_3^2-k_2^2)t+c_2,\,\xi_3=k_3\eta_3+c_3,$$
$$\xi_2+\xi_3=(k_2+k_3)\eta_1+\frac{k_3p_2-k_2p_3}{k_3}y+k_2(k_3^2-k_2^2)t+c_4.$$

{\bf (a)} In the case of $t\to +\infty$: We have $\xi_3\approx c,\,\xi_1\approx c,\,\xi_3\to -\infty,\,\xi_2+\xi_3\to -\infty$, and then Eq.\ \eqref{3soliton1} becomes $f\sim 1+\e^{\xi_1}+\e^{\xi_3}$. This situation is the same as the case {\bf I (a)} above.

{\bf (b)} In the case of $t\to -\infty,\,y\to -\infty$: We have $\xi_3\approx c,\,\xi_1\to -\infty,\,\xi_3\to +\infty,\,\xi_2+\xi_3\to +\infty$.  and then Eq.\ \eqref{3soliton1} becomes
$$f\sim \e^{\xi_3}+a_{23}\e^{\xi_2+\xi_3},\,t\to -\infty,\,y\to -\infty.$$

Some of the asymptotic forms obtained above are repeated. After sorting, we get the asymptotic forms of the four arms as follows,

Before collision ($t\to -\infty$):
\begin{flalign}\label{asyf01}
\begin{split}
&f_2^-\sim 1+\e^{\eta_2},\,f_3^-\sim 1+a_{23}\e^{\eta_3},\,(y\to -\infty),\\
&f_1\sim 1+\e^{\eta_1},\,f_{1-2-3}\sim 1+a_{23}\e^{\eta_2+\eta_3-\eta_1},\,(y\to +\infty).
\end{split}
\end{flalign}

After collision ($t\to +\infty$):
\begin{flalign}\label{asyf02}
\begin{split}
&f_2^+\sim 1+a_{23}\e^{\eta_2},\,f_3^+\sim 1+\e^{\eta_3},\,(y\to -\infty),\\
&f_1\sim 1+\e^{\eta_1},\,f_{1-2-3}\sim 1+a_{23}\e^{\eta_2+\eta_3-\eta_1},\,(y\to +\infty).
\end{split}
\end{flalign}

\begin{rk}
Notations $^{+}$ and $^-$ are used to distinguish between $f$ before and after the interaction. Comparing Eq.\ \eqref{asyf01} and Eq.\ \eqref{asyf02}, $f_j^{+}$ and $f_j^{-}$ differ by a phase shift at $t=\pm \infty$.
\end{rk}

\begin{rk}
Eq.\ \eqref{asy00} is the exactly same as Eq.\ \eqref{asyf01} for $t\rightarrow \infty$, which means that if we use the asymptotic analysis method in Ref.\ \cite{jpsj1983-1, pla2023,rao01,prsa2021} (consider $t$ to be constant), we cannot get the asymptotic form \eqref{asyf02} for $t\rightarrow +\infty$. The main difference
between Eq.\ \eqref{asyf01} and Eq.\ \eqref{asyf02}, as we have mentioned below Eq.\ \eqref{asy00}, is due to the different appearance of phase shift term $a_{23}$, although both of them have four soliton arms.
\end{rk}

Further, we can get the following proposition,
\begin{prop}\label{prop3.1}
The asymptotic forms of the weak 2-resonant 3-soliton with $0<k_1=k_3<k_2,\,p_1=p_2>p_3>0$ are as following:

Before collision ($t\to-\infty$):
\begin{flalign}\label{asy01}
\begin{split}
y\to +\infty,\,S_1:\qquad&u_1\approx -\frac{k_1p_1}{2}\sech^2(\frac{\xi_1}{2}),\,v_1\approx -\frac{k_1^2}{2}\sech^2(\frac{\xi_1}{2}),\\
S_{1-2-3}:\,\,&\widehat{u_{1-2-3}}\approx -\frac{(k_1- k_2- k_3)(p_1- p_2- p_3)}{2}\sech^2(\frac{\xi_1- \xi_2- \xi_3- \ln a_{23}}{2}),\\
&\widehat{v_{1-2-3}}\approx -\frac{(k_1-k_2- k_3)^2}{2}\sech^2(\frac{\xi_1- \xi_2- \xi_3- \ln a_{23}}{2}).\\
y\to -\infty,\,S_2:\qquad&u_2\approx -\frac{k_2p_2}{2}\sech^2(\frac{\xi_2}{2}),\,v_2\approx -\frac{k_2^2}{2}\sech^2(\frac{\xi_2}{2}),\\
S_3:\qquad&\widehat{u_3}\approx -\frac{k_3p_3}{2}\sech^2(\frac{\xi_3+\ln a_{23}}{2}),\,\widehat{v_3}\approx -\frac{k_3^2}{2}\sech^2(\frac{\xi_3}{2}).
\end{split}
\end{flalign}

After collision ($t\to+\infty$):
\begin{flalign}\label{asy02}
\begin{split}
y\to +\infty,\,S_1:\qquad &u_1\approx -\frac{k_1p_1}{2}\sech^2(\frac{\xi_1}{2}),\,v_1\approx -\frac{k_1^2}{2}\sech^2(\frac{\xi_1}{2}),\\
S_{1-2-3}:\,\,&\widehat{u_{1-2-3}}\approx -\frac{(k_1- k_2- k_3)(p_1- p_2- p_3)}{2}\sech^2(\frac{\xi_1- \xi_2- \xi_3- \ln a_{23}}{2}),\\
&\widehat{v_{1-2-3}}\approx -\frac{(k_1-k_2- k_3)^2}{2}\sech^2(\frac{\xi_1- \xi_2- \xi_3- \ln a_{23}}{2}).\\
y\to -\infty,\,S_2:\qquad&\widehat{u_2}\approx -\frac{k_2p_2}{2}\sech^2(\frac{\xi_2+\ln a_{23}}{2}),\,\widehat{v_2}\approx -\frac{k_2^2}{2}\sech^2(\frac{\xi_2}{2}),\\
S_3:\qquad&u_3\approx -\frac{k_3p_3}{2}\sech^2(\frac{\xi_3}{2}),\,v_3\approx -\frac{k_3^2}{2}\sech^2(\frac{\xi_3}{2}).
\end{split}
\end{flalign}
\end{prop}

\begin{rk}
In this paper, $S_j$ corresponds to the soliton arms determined by $f_j^-,\,f_j^+$ and $f_j$. In contrast with Proposition \ref{prop3.1}, we find that analyzing the asymptotic form along the region $\zeta_j = x + \frac{p_j}{k_j}y = \text{constant}$ while considering $t$ as a constant yields incomplete results (see Eq.\ \eqref{asy00}). Therefore, in this paper, we utilize the same method as Proposition \ref{prop3.1} for our analysis.
\end{rk}
\begin{rk}
To get the asymptotic forms \eqref{asy01} and \eqref{asy02}, we need to make sure that $k_2p_3-k_3p_2>0$. The same is true of the asymptotic forms \eqref{asy03} and \eqref{asy04} in section 3.3.
\end{rk}
\begin{rk}
It can be seen from \eqref{asy01} and \eqref{asy02} that the asymptotic forms of the four arms are partially changed to $t$: The asymptotic form of $S_2$ and $S_3$ differ by a phase shift $\ln a_{23}$ at $t\to-\infty$ and $t \to +\infty$. If $a_{23}=1$, the asymptotic forms \eqref{asy01} and \eqref{asy02} are the same, and the four arms shift over time without a phase shift. The same is true of the asymptotic forms \eqref{asy03} and \eqref{asy04} in section 3.3.
\end{rk}

Especially, there are a special asymptotic form $f\sim \e^{\xi_1}+\e^{\xi_3}$ with $t\to+\infty,\,x\to \pm \infty$. It is generated by the resonances between $S_1$ and $S_3$. Simultaneously, $S_{1-3}$ (corresponding to $f\sim \e^{\xi_1}+\e^{\xi_3}$) resonates with $S_2$, leading to the emergence of $S_{1-2-3}$. Considering the combined effects of these two sets of resonances and the location of the four arms ($S_1,\,S_2,\,S_3$ and $S_{1-2-3}$), we find the arms $S_{1-3}$ manifests as a spatially finite structure, which is the stem structure we will study in this paper. And then, we investigate the behaviour of the intermediate regions for $t\to-\infty$. The resonance between soliton $S_1$ and soliton $S_2$ produces $S_{1-2}$ (corresponding to $f\sim \e^{\xi_1}+\e^{\xi_2}$); simultaneously, $S_{1-2}$ resonates with $S_3$, resulting in the $S_{1-2-3}$. Due to the combined effects of these two sets of resonances and considering the location of the four arms, we find the arms $S_{1-3}$ manifests as a spatially finite structure. Then, the asymptotic forms of the stems are given by
\begin{flalign}
\begin{split}
f_{1-2}\sim \e^{\eta_1}+\e^{\eta_2},\,t\to -\infty,\,\text{and}\,f_{1-3}\sim \e^{\eta_1}+\e^{\eta_3},\,t\to +\infty.
\end{split}
\end{flalign}

It reveals a distinctive feature of the weak 2-resonant 3-soliton: a completely different stem structure with variable length before and after collision ($t\to\pm\infty$). Naturally, we have the following proposition,
\begin{prop}\label{prop3.2}
The stem structures corresponding to asymptotic forms \eqref{asy01} and \eqref{asy02} are as follows:
\begin{flalign}\label{stem01}
\begin{split}
&S_{1-2}:\,u_{1-2}\approx -\frac{(k_1-k_2)(p_1-p_2)}{2}\sech^2(\frac{\xi_1-\xi_2}{2}),\,v_{1-2}\approx -\frac{(k_1-k_2)^2}{2}\sech^2(\frac{\xi_1-\xi_2}{2}),\,t\to-\infty.\\
&S_{1-3}:\,u_{1-3}\approx -\frac{(k_1-k_3)(p_1-p_3)}{2}\sech^2(\frac{\xi_1-\xi_3}{2}),\,v_{1-3}\approx -\frac{(k_1-k_3)^2}{2}\sech^2(\frac{\xi_1-\xi_3}{2}),\,t\to+\infty.
\end{split}
\end{flalign}
\end{prop}

\begin{figure}[h!tb]
	\centering
\subfigure[$u:\,t=-8$]{\includegraphics[height=3.4cm,width=3.4cm]{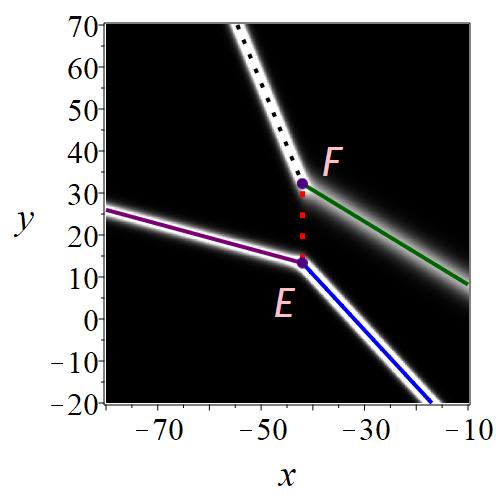}}
    \subfigure[$u:\,t=-4$]{\includegraphics[height=3.4cm,width=3.4cm]{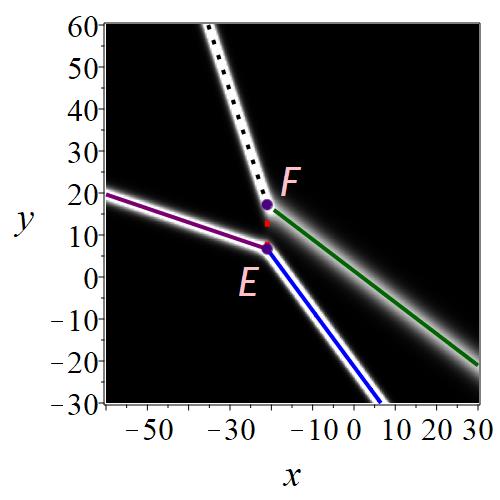}}
	\subfigure[$u:\,t=0$]{\includegraphics[height=3.4cm,width=3.4cm]{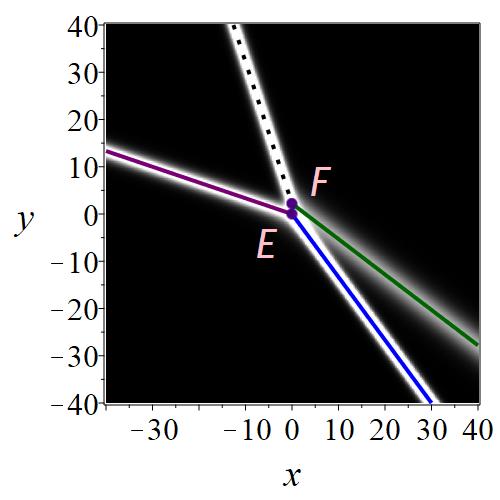}}
	\subfigure[$u:\,t=5$]{\includegraphics[height=3.4cm,width=3.4cm]{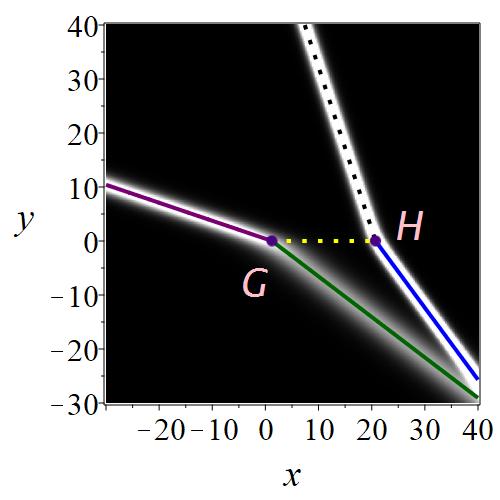}}
    \subfigure[$u:\,t=10$]{\includegraphics[height=3.4cm,width=3.4cm]{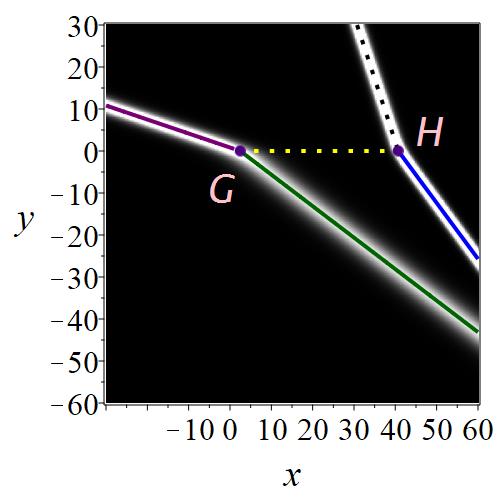}}\\
	\subfigure[$v:\,t=-8$]{\includegraphics[height=3.4cm,width=3.4cm]{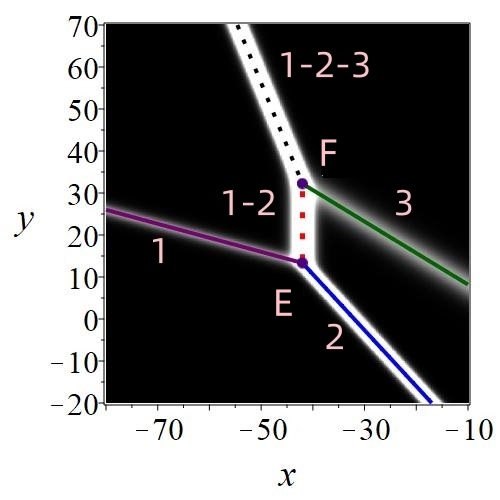}}
    \subfigure[$v:\,t=-4$]{\includegraphics[height=3.4cm,width=3.4cm]{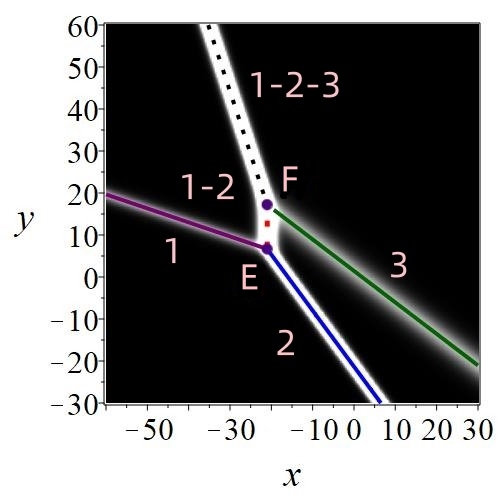}}
	\subfigure[$v:\,t=0$]{\includegraphics[height=3.4cm,width=3.4cm]{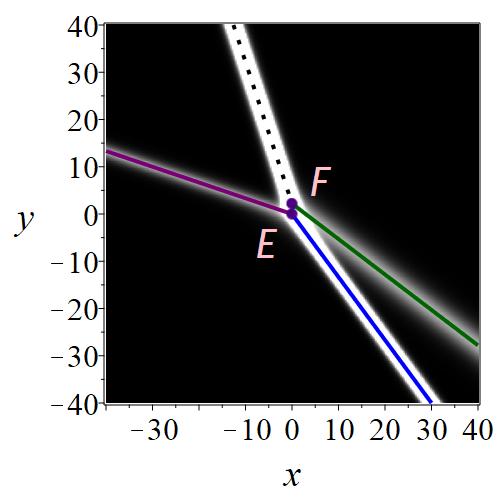}}
	\subfigure[$v:\,t=5$]{\includegraphics[height=3.4cm,width=3.4cm]{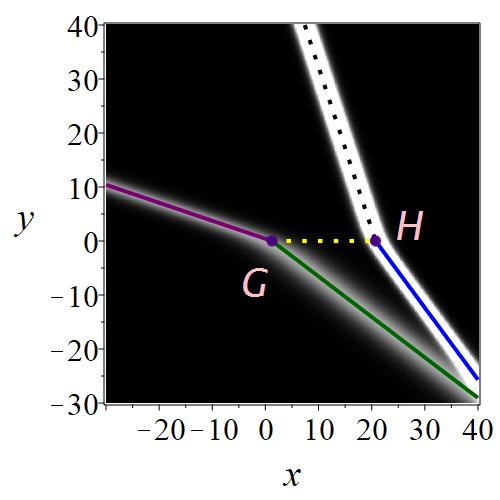}}
  \subfigure[$v:\,t=10$]{\includegraphics[height=3.4cm,width=3.4cm]{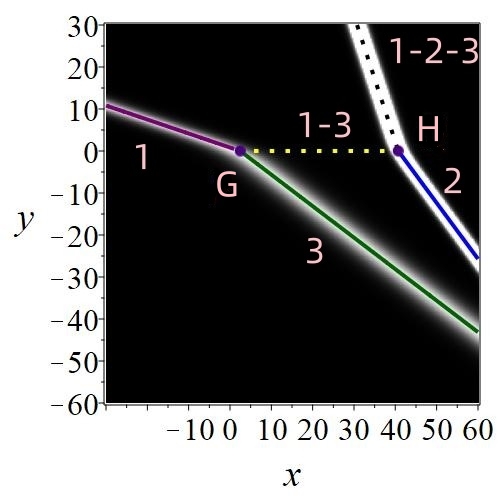}}
	\caption{The density plots of the weak 2-resonant 3-soliton given by \eqref{uv} and \eqref{3soliton1} with 
  $k_1=\frac{1}{2},\,k_2=2,\,k_3=\frac{1}{2},\,p_1=\frac{3}{2},\,p_2=\frac{3}{2},\,p_3=\frac{2}{3},\,\xi_1^0=0,\,\xi_2^0=0,\,\xi_3^0=0$. 
  The lines are the trajectories of the arms and stem structures, and the points are the endpoints of the variable length stem structures.
	}\label{fig3-1}
\end{figure}

The asymptotic forms given by proposition \ref{prop3.1} and \ref{prop3.2} reveal that the 2-resonant 3-soliton has four arms and a stem structure, and its evolution with time is illustrated in Fig.\ \ref{fig3-1}. When $t\to -\infty$, the stem structure $S_{1-2}$ connects two pairs of V-shape solitons ($S_1$ and $S_2$, $S_1$ and $S_{1-2-3}$). As time evolves, the length of this stem structure gradually diminishes until it vanishes around $t=0$. At this moment, the four arms ($S_1,\,S_2,\,S_3,\,S_{1-2-3}$) intersect together, and the two pairs of V-shape solitons turn into $S_1$ and $S_3$, $S_2$ and $S_{1-2-3}$. As time passes ($t\to +\infty$), a new stem structure $S_{1-3}$ appears, gradually growing longer and connecting these two pares of V-shape solitons. This phenomenon is called soliton reconnection.

\begin{table}[http]
  \centering
  \caption{Physical quantities of the arms with $0<k_1=k_3<k_2,\,p_1=p_2>p_3>0$}
  \label{tab:t2}
  \begin{tabular}{cccccc}
    \Xhline{1pt}
     Soliton & Trajectory & Velocity & Amplitude & Components\\
    \hline
    \multirow{2}{*}{$S_{1-2}$} & \multirow{2}{*}{$l_{1-2}$} & \multirow{2}{*}{$(k_1^2+k_1k_2+k_2^2,\,0)$} & $0$ & $u_{1-2}$ \\
     & & &$-\frac{(k_1-k_2)^2}{2}$ & $v_{1-2}$ \\
    \hline
    \multirow{2}{*}{$S_{1-3}$} & \multirow{2}{*}{$l_{1-3}$} & \multirow{2}{*}{$(0,\,0)$} & $0$ & $u_{1-3}$ \\
     & & &$0$ & $v_{1-3}$ \\
    \hline
    \multirow{2}{*}{$S_{1-2-3}$} & \multirow{2}{*}{$\widehat{l_{1-2-3}}$} & \multirow{2}{*}{$(k_2^2,\,-\frac{k_2^3}{p_3})$} & $\frac{k_2p_3}{2}$ & $\widehat{u_{1-2-3}}$ \\
     & & &$-\frac{k_2^2}{2}$ & $\widehat{v_{1-2-3}}$ \\
    \Xhline{1pt}
  \end{tabular}
   \caption*{\captionsetup{justification=raggedright,singlelinecheck=false,format=hang} \quad The solitons $S_j$ $(j=1-2,\,1-3,\,1-2-3)$ are composed by two components $u_{j}$ and $v_{j}$, and their trajectories are listed by \eqref{l01}.}
\end{table}

\subsection{The stem structure with $0<k_1=k_3<k_2,\,p_1=p_2>p_3>0$}
The trajectories, amplitudes, and velocities of the four arms and two stems are listed in Tables \ref{tab:t1} and \ref{tab:t2}, where
\begin{flalign}\label{l01}
\begin{split}
&\boldsymbol{l_1:}\,\xi_1=0,\,\quad\,\boldsymbol{l_2:}\, \xi_2=0,\quad\,\boldsymbol{l_3:}\, \xi_3=0,\,\quad \boldsymbol{\widehat{l_2}:}\,\xi_2+\ln a_{23}=0,\,\quad \boldsymbol{\widehat{l_3}:}\,\xi_3+\ln a_{23}=0,\\
&\boldsymbol{l_{1-2}:}\, \xi_1-\xi_2=0,\,\quad \boldsymbol{l_{1-3}:}\,\xi_1-\xi_3=0,\,\quad \boldsymbol{\widehat{l_{1-2-3}}:}\,\xi_1-\xi_2-\xi_3-\ln a_{23}=0.
\end{split}
\end{flalign}
Here, $\widehat{l_j}$ means that the formula of the trajectory contains $a_{23}$, otherwise not included, and its corresponding asymptotic form is $\widehat{u_j}$ and $\widehat{v_j}$.

It's not hard to prove that the trajectories of $S_1$, $S_2$, $S_{1-2}$ intersects at a point, and that the trajectories of $S_3$, $S_{1-2-3}$, $S_{1-2}$ intersect at a point at $t\to -\infty$. In the same way, the trajectories of $S_1$, $S_3$, $S_{1-3}$ intersects at a point, and that the trajectories of $S_2$, $S_{1-2-3}$, $S_{1-3}$ intersect at a point at $t\to +\infty$. This further confirmed the correctness of the location of the stem structures \eqref{stem01}. Notably, the amplitudes of $v_{1-2},\,u_{1-2},\,u_{1-3}$ all equal zero, implying that $u$ effectively exhibits only four arms and a degenerated stem structure. It can be observed that $l_{1-2}$ is oriented vertically, while $l_{1-3}$ is oriented horizontally. Consequently, when $t\to -\infty$, the vertical variable length stem structure $S_{1-2}$, generated by resonance, gradually diminishes with time increase in length until it dissipates around $t=0$. The resulting horizontal soliton exhibits an amplitude of zero. Interestingly, the trajectories and velocities of $u$ and $v$ are the same. Fig.\ \ref{fig3-1} illustrates the trajectories of $u$ and $v$ at various moments, with the background representing the density plot. By solving for the intersection points of these trajectories, one can ascertain the endpoints of the variable-length stem structures as follows,
\begin{flalign}\label{EFGH}
\begin{split}
&E\,\left((k_1^2+k_1k_2+k_2^2)t-\frac{\xi_1^0-\xi_2^0}{k_1-k_2},\,-\frac{k_1k_2(k_1+k_2)t}{p_1}+\frac{k_2\xi_1^0-k_1\xi_2^0}{p_1(k_1-k_2)}\right),\,\\
&F\,\left((k_1^2+k_1k_2+k_2^2)t-\frac{\xi_1^0-\xi_2^0}{k_1-k_2},\,-\frac{k_1k_2(k_1+k_2)t+\ln a_{23}+\xi_3^0}{p_3}+\frac{k_1(\xi_1^0-\xi_2^0)}{p_3(k_1-k_2)}\right),\,\\
&G\,\left(k_1^2t-\frac{p_1\xi_3^0-p_3\xi_1^0}{k_1(p_1-p_3)},\,-\frac{\xi_1^0-\xi_3^0}{p_1-p_3}\right),\,H\,\left(k_2^2t-\frac{\ln a_{23}}{k_2}+\frac{p_1(\xi_1^0-\xi_2^0-\xi_3^0)+p_3\xi_2^0}{k_2(p_1-p_3)},\,-\frac{\xi_1^0-\xi_3^0}{p_1-p_3}\right).
\end{split}
\end{flalign}

In this context, the point $E$ corresponds to the intersection of lines $l_1$ and $l_2$, while point $F$ designates the intersection of lines $\widehat{l_3}$ and $\widehat{l_{1-2-3}}$. 
Point $G$ marks the intersection of lines $l_1$ and $l_3$, and point $H$ denotes the intersection of lines $\widehat{l_2}$ and $\widehat{l_{1-2-3}}$, respectively. These points are also shown in Fig.\ \ref{fig3-1}. 
Whereupon, the lengths of the trajectories of the variable length stem structures are obtained as follows,
{\small
\begin{flalign}\label{EF+GH}
\begin{split}
&|EF|=\left|\frac{k_1k_2(p_1-p_3)(k_1^2-k_2^2)t+p_1(k_1-k_2)\ln a_{23}-k_1p_1(\xi_1^0-\xi_2^0-\xi_3^0)-k_1p_3\xi_2^0-k_2p_1\xi_3^0+k_2p_3\xi_1^0}{p_1p_3(k_1-k_2)}\right|,\\
&|GH|=\left|\frac{k_1k_2(p_1-p_3)(k_1^2-k_2^2)t+k_1(p_1-p_3)\ln a_{23}-k_1p_1(\xi_1^0-\xi_2^0-\xi_3^0)-k_1p_3\xi_2^0-k_2p_1\xi_3^0+k_2p_3\xi_1^0}{k_1k_2(p_1-p_3)}\right|.
\end{split}
\end{flalign}}

\begin{rk}
Given the intricate complexity of the evolution around $t=0$, the validity of the Eqs.\ \eqref{EFGH} and \eqref{EF+GH} is constrained to scenarios where $t\ll0$ and $t\gg0$.
This constraint equally applies to the subsequent Eqs.\ \eqref{PQMN} and \eqref{PQ+MN}.
\end{rk}

\begin{figure}[h!tb]
	\centering
	\subfigure[$v|_{l_{1-2}}^{(1)}:\,t=-8$]{\includegraphics[height=3.4cm,width=3.4cm]{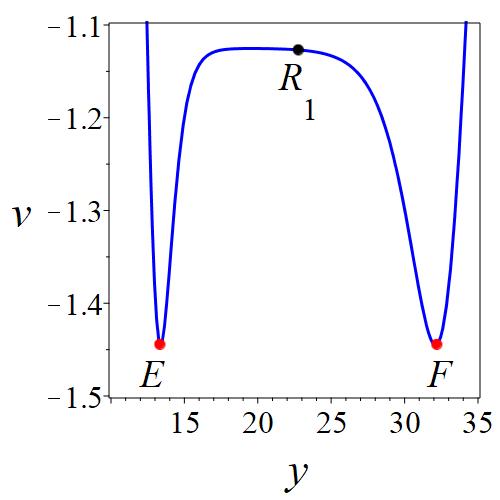}}
    \subfigure[$v|_{l_{1-3}}^{(1)}:\,t=5$]{\includegraphics[height=3.4cm,width=3.4cm]{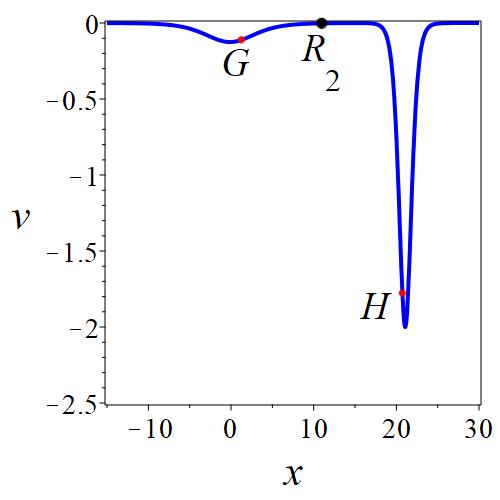}}
    \subfigure[$v|_{l_{1-2}}^{(2)}:\,t=-8$]{\includegraphics[height=3.4cm,width=3.4cm]{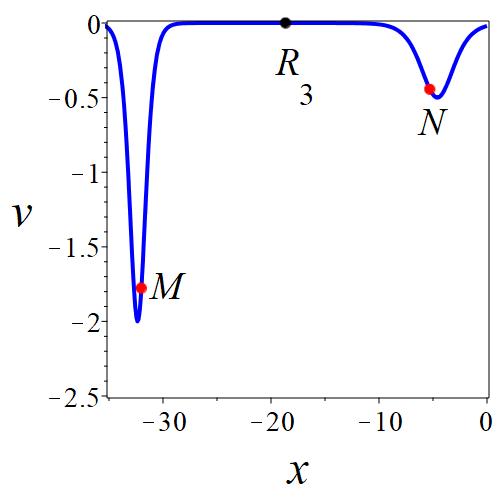}}
    \subfigure[$v|_{l_{1-3}}^{(2)}:\,t=5$]{\includegraphics[height=3.4cm,width=3.4cm]{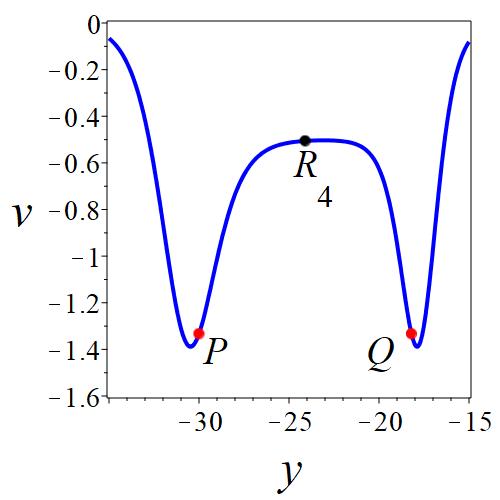}}
	\caption{(a) (b) The cross-sectional curves \eqref{cross31};
   (c) (d) The cross-sectional curves \eqref{cross32}. 
   The red points are the endpoints of the variable length stem structures.}
	\label{fig3-4}
\end{figure}

Now we will analyze the amplitudes of variable length stem structures next. 
Since the correlation analysis methods for $u$ and $v$ are identical, we only analyze $v$ here. 
Below, we present only the formulas related to $v$, and the formulas for $u$ are provided in the appendix. 
For computational convenience, we set $k_1=\frac{1}{2},\,k_2=2,\,k_3=\frac{1}{2},\,p_1=\frac{3}{2},\,p_2=\frac{3}{2},\,p_3=\frac{2}{3},\,\xi_1^0=0,\,\xi_2^0=0,\,\xi_3^0=0$.
The cross-sectional curves of 3-soliton \eqref{uv} with Eqs.\ \eqref{3soliton1} along $l_{1-2}$ and $l_{1-3}$ shown in Fig.\ \ref{fig3-4} (a) and (b) are expressed as,
\begin{flalign}\label{cross31}
\begin{split}
&v|_{l_{1-2}}^{(1)}=-\frac{13}{2}\cdot\frac{51\e^{\frac{15t}{2}+\frac{11y}{3}}+48\e^{\frac{15t}{2}+\frac{17y}{6}}+117\e^{5t+3y}+192\e^{5t+\frac{13y}{6}}+221\e^{\frac{5t}{2}+\frac{3y}{2}}
+13\e^{\frac{5t}{2}+\frac{2y}{3}}}{(3\e^{5t+\frac{13y}{6}}+26\e^{\frac{5t}{2}+\frac{3y}{2}}+13\e^{\frac{5t}{2}+\frac{2y}{3}}+13)^2},\\
&v|_{l_{1-3}}^{(1)}=-\frac{13}{2}\cdot\frac{96\e^{3x-\frac{33t}{4}}+3\e^{\frac{9x}{2}-\frac{129t}{8}}+309\e^{\frac{5x}{2}-\frac{65t}{8}}+26\e^{\frac{x}{2}-\frac{t}{8}}+208\e^{2x-8t}}
{(3\e^{\frac{5x}{2}-\frac{65t}{8}}+26\e^{\frac{x}{2}-\frac{t}{8}}+13\e^{2x-8t}+13)^2}.
\end{split}
\end{flalign}

\begin{figure}[h!tb]
	\centering
    \raisebox{15ex}{(a)}\includegraphics[height=3.8cm,width=3.5cm]{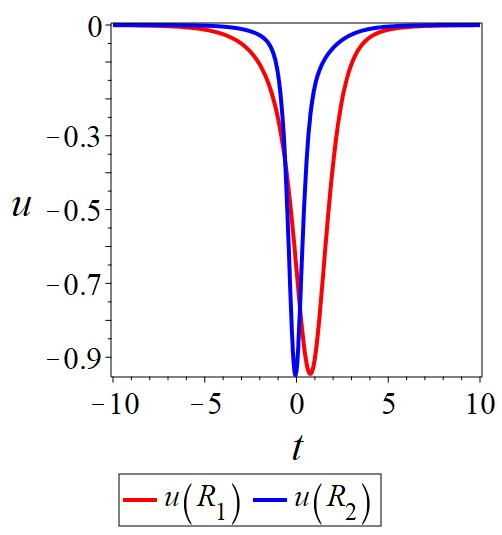}
    \raisebox{15ex}{(b)}\includegraphics[height=3.8cm,width=3.5cm]{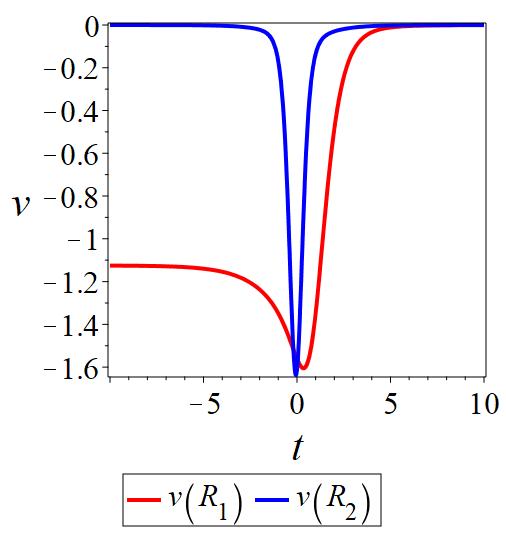}
    \raisebox{15ex}{(c)}\includegraphics[height=3.8cm,width=3.5cm]{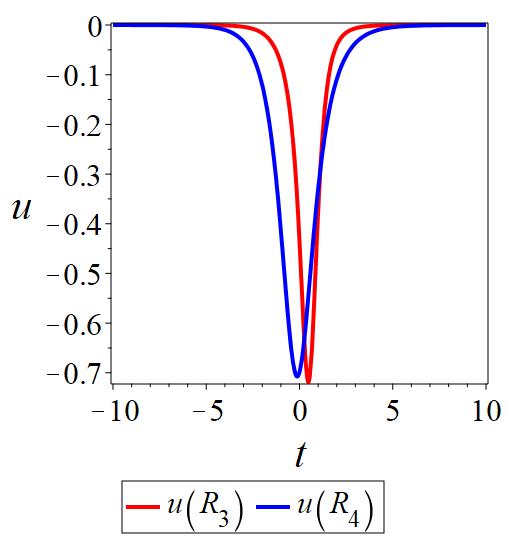}
    \raisebox{15ex}{(d)}\includegraphics[height=3.8cm,width=3.5cm]{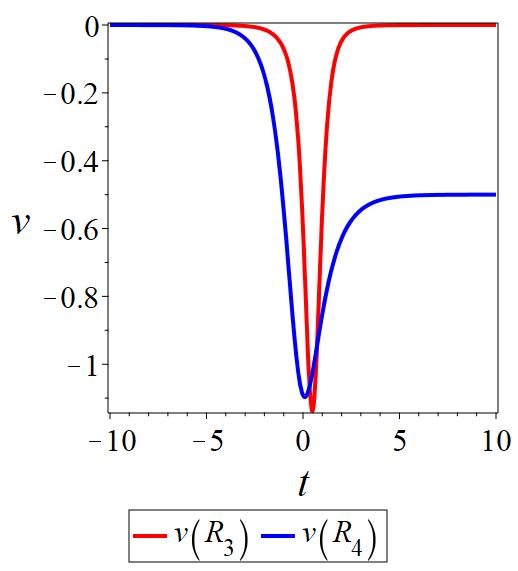}
	\caption{The evolutions of the amputation on the point $R_j$ over time: 
 (a)--(d) correspond to \eqref{amplocaladd1} \eqref{amplocal1} \eqref{amplocaladd2}\eqref{amplocal2}, respectively.	
 }\label{fig3-3}
\end{figure}

Due to the intricacy of the calculations, providing explicit expressions for the extreme points along $l_{1-2}$ and $l_{1-3}$ is difficult. 
To approximate the amplitude of the variable length stem structures, we focus on the amplitudes at the midpoint of $EF$ and $GH$, which are denoted as $R_3$ and $R_4$ respectively, to approximate the amplitudes of $S_{1-2}$ and $S_{1-3}$. 
In instances where $|t|\gg 0$, the amplitudes of both $u_{1-2}$ and $u_{1-3}$ are rendered as zero. 
Consequently, we exclusively present the amplitude trend plot for component $u$ in Fig.\ \ref{fig3-3}, while for component $v$, we provide both the amplitude formula and the corresponding trend plot across time. 
The amplitudes of $S_{1-2}$ and $S_{1-3}$ are expressed as:
\begin{flalign}\label{amplocal1}
\begin{split}
S_{1-2}:\, &v(R_1)=-\frac{208\alpha_1\e^{\frac{425t}{144}}+221\alpha_1\e^{\frac{25t}{16}}+192\alpha_5
\e^{\frac{325t}{144}}+221\alpha_2\e^{\frac{25t}{36}}+3\e^{\frac{275t}{72}}-507\alpha_2}{2\sqrt{39}
\Big(\alpha_3\e^{\frac{425t}{144}}+\e^{\frac{325t}{144}}+\frac{3}{\sqrt{39}}\e^{\frac{25t}{16}}+\frac{26\alpha_1}{3}\Big)^2},\,\\
S_{1-3}:\, &v(R_2)=-\frac{3744\alpha_2\e^{\frac{45t}{8}}+1205\alpha_4\e^{\frac{75t}{16}}+
8112\e^{\frac{15t}{4}}+1014\alpha_1\e^{\frac{135t}{16}}+507\alpha_1\e^{\frac{15t}{16}}}
{2\sqrt{39}\left(26\alpha_1\e^{\frac{75t}{16}}+3\alpha_4\e^{\frac{15t}{16}}+\sqrt{39}\e^{\frac{15}{4}}+13\right)^2},
\end{split}
\end{flalign}
where $\alpha_1=\left(\frac{3}{13}\right)^{\frac{3}{8}},\,\alpha_2=\left(\frac{13}{3}\right)^{\frac{1}{4}},\,
\alpha_3=\left(\frac{3}{13}\right)^{\frac{1}{8}},\,\alpha_4=\left(\frac{13}{3}\right)^{\frac{1}{8}},\,\alpha_5=\sqrt[8]{39}$.

Just check out the limits:
\[
\lim_{t\to +\infty} v(R_1) = 0, \quad \lim_{t\to -\infty} v(R_1) = 0, \quad \lim_{t\to +\infty} v(R_2) = -\frac{9}{8}, \quad \lim_{t\to -\infty} v(R_2) = 0.
\]
You can also see this in Fig.\ \ref{fig3-3}. 
The amplitudes change with time are shown in Fig.\ \ref{fig3-3} (c). 
As can be seen from the figure that $S_{1-2}$ disappears and $S_{1-3}$ arises around $t=0$.
\begin{rk}
In this section, $R_j$ is the midpoint of the variable length stem structures. 
Because the exact amplitude (the extreme value of the amplitude of the  variable length stem structure) is difficult to solve analytically, we use $v(R_j)$ as the approximate amplitude of the variable length stem structure.
\end{rk}

\subsection{The asymptotic analysis and stem structures with $k_1=k_2> k_3>0,\,0<p_1=p_3< p_2$}
Using the same method as in the section 3.1, the following propositions can be obtained.
\begin{prop}\label{prop3.3}
The asymptotic forms of the weak 2-resonant 3-soliton with $k_1=k_2> k_3>0,\,0<p_1=p_3< p_2$ are as following:

Before collision ($t\to-\infty$):
\begin{flalign}\label{asy03}
\begin{split}
y\to -\infty,\,S_1:\qquad&u_1\approx -\frac{k_1p_1}{2}\sech^2(\frac{\xi_1}{2}),\,v_1\approx -\frac{k_1^2}{2}\sech^2(\frac{\xi_1}{2}),\\
S_{1-2-3}:\,\,&\widehat{u_{1-2-3}}\approx -\frac{(k_1- k_2- k_3)(p_1- p_2- p_3)}{2}\sech^2(\frac{\xi_1- \xi_2- \xi_3- \ln a_{23}}{2}),\\
&\widehat{v_{1-2-3}}\approx -\frac{(k_1-k_2- k_3)^2}{2}\sech^2(\frac{\xi_1- \xi_2- \xi_3- \ln a_{23}}{2}).\\
y\to +\infty,\,S_2:\qquad&u_2\approx -\frac{k_2p_2}{2}\sech^2(\frac{\xi_2}{2}),\,v_2\approx -\frac{k_2^2}{2}\sech^2(\frac{\xi_2}{2}),\\
S_3:\qquad&\widehat{u_3}\approx -\frac{k_3p_3}{2}\sech^2(\frac{\xi_3+\ln a_{23}}{2}),\,\widehat{v_3}\approx -\frac{k_3^2}{2}\sech^2(\frac{\xi_3}{2}).
\end{split}
\end{flalign}

After collision ($t\to+\infty$):
\begin{flalign}\label{asy04}
\begin{split}
y\to -\infty,\,S_1:\qquad &u_1\approx -\frac{k_1p_1}{2}\sech^2(\frac{\xi_1}{2}),\,v_1\approx -\frac{k_1^2}{2}\sech^2(\frac{\xi_1}{2}),\\
S_{1-2-3}:\,\,&\widehat{u_{1-2-3}}\approx -\frac{(k_1- k_2- k_3)(p_1- p_2- p_3)}{2}\sech^2(\frac{\xi_1- \xi_2- \xi_3- \ln a_{23}}{2}),\\
&\widehat{v_{1-2-3}}\approx -\frac{(k_1-k_2- k_3)^2}{2}\sech^2(\frac{\xi_1- \xi_2- \xi_3- \ln a_{23}}{2}).\\
y\to +\infty,\,S_2:\qquad&\widehat{u_2}\approx -\frac{k_2p_2}{2}\sech^2(\frac{\xi_2+\ln a_{23}}{2}),\,\widehat{v_2}\approx -\frac{k_2^2}{2}\sech^2(\frac{\xi_2}{2}),\\
S_3:\qquad&u_3\approx -\frac{k_3p_3}{2}\sech^2(\frac{\xi_3}{2}),\,v_3\approx -\frac{k_3^2}{2}\sech^2(\frac{\xi_3}{2}).
\end{split}
\end{flalign}
\end{prop}

\begin{prop}\label{prop3.4}
The stem structures corresponding to asymptotic forms \eqref{asy03} and \eqref{asy04} are as following:
\begin{flalign}\label{stem02}
\begin{split}
&S_{1-2}: \, u_{1-2}\approx -\frac{(k_1-k_2)(p_1-p_2)}{2}\sech^2(\frac{\xi_1-\xi_2}{2}),\,v_{1-2}\approx -\frac{(k_1-k_2)^2}{2}\sech^2(\frac{\xi_1-\xi_2}{2}),\,t\to-\infty.\\
&S_{1-3}:\, u_{1-3}\approx -\frac{(k_1-k_3)(p_1-p_3)}{2}\sech^2(\frac{\xi_1-\xi_3}{2}),\,v_{1-3}\approx -\frac{(k_1-k_3)^2}{2}\sech^2(\frac{\xi_1-\xi_3}{2}),\,t\to+\infty.
\end{split}
\end{flalign}
\end{prop}

\begin{figure}[h!tb]
	\centering
 \subfigure[$u:\,t=-8$]{\includegraphics[height=3.4cm,width=3.4cm]{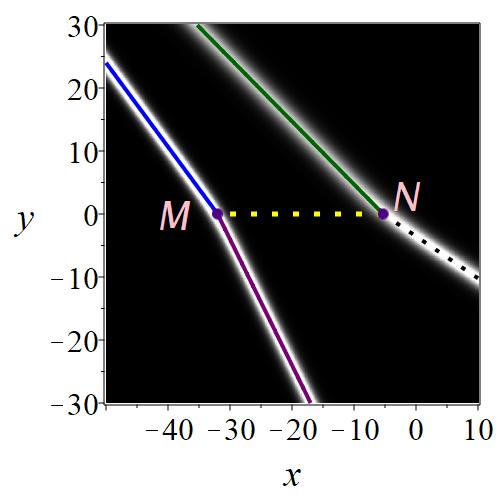}}
    \subfigure[$u:\,t=-4$]{\includegraphics[height=3.4cm,width=3.4cm]{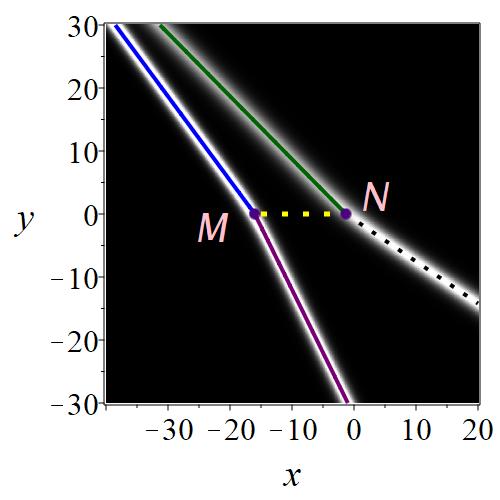}}
	\subfigure[$u:\,t=0$]{\includegraphics[height=3.4cm,width=3.4cm]{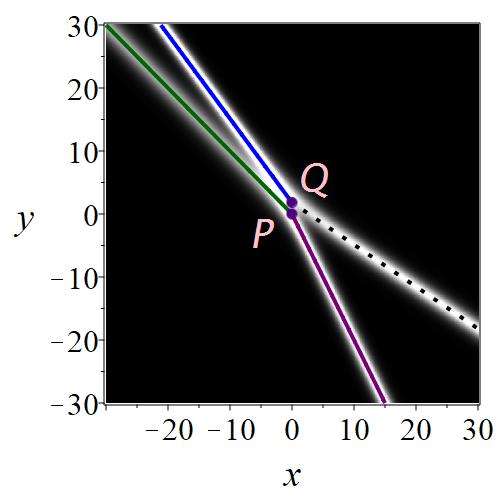}}
	\subfigure[$u:\,t=5$]{\includegraphics[height=3.4cm,width=3.4cm]{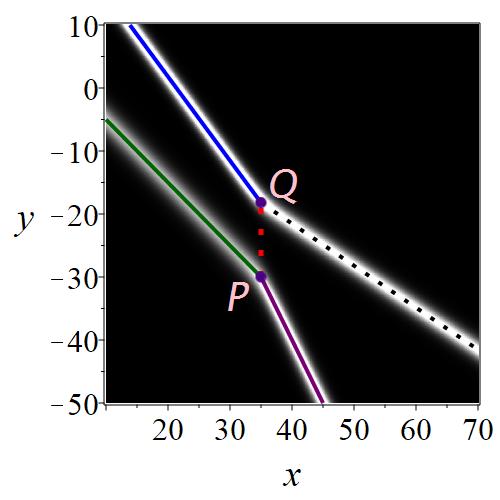}}
    \subfigure[$u:\,t=10$]{\includegraphics[height=3.4cm,width=3.4cm]{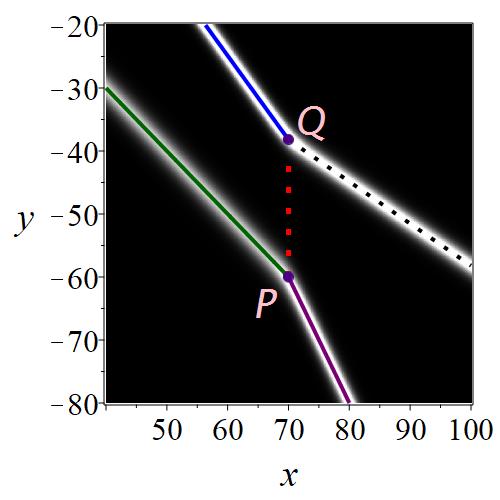}}\\
    \subfigure[$v:\,t=-8$]{\includegraphics[height=3.4cm,width=3.4cm]{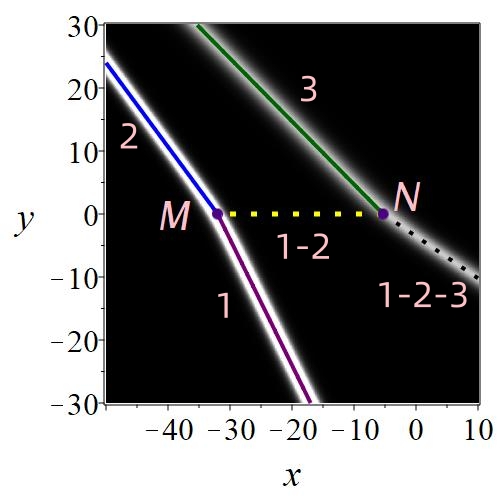}}
    \subfigure[$v:\,t=-4$]{\includegraphics[height=3.4cm,width=3.4cm]{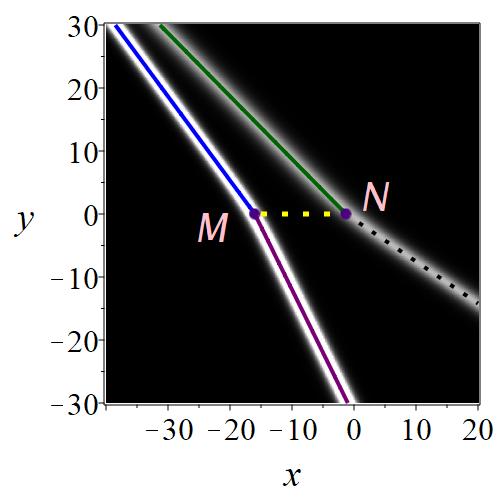}}
	\subfigure[$v:\,t=0$]{\includegraphics[height=3.4cm,width=3.4cm]{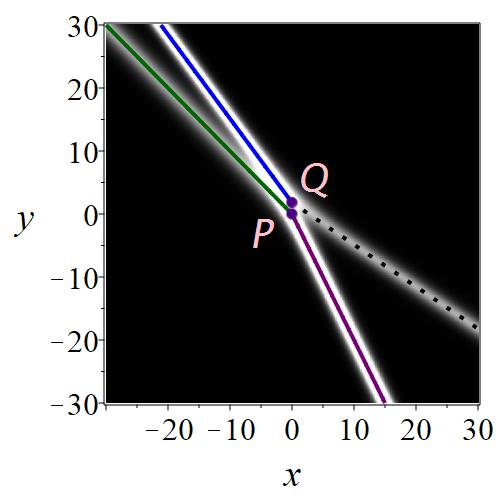}}
	\subfigure[$v:\,t=5$]{\includegraphics[height=3.4cm,width=3.4cm]{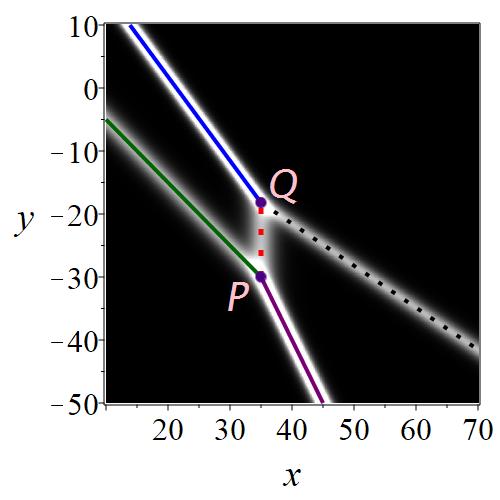}}
    \subfigure[$v:\,t=10$]{\includegraphics[height=3.4cm,width=3.4cm]{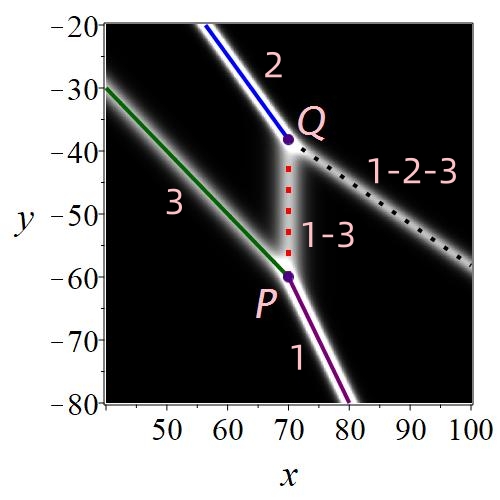}}
	\caption{The density plots of the weak 2-resonant 3-soliton given by \eqref{uv} and \eqref{3soliton1} with 
  $k_1=2,\,k_2=2,\,k_3=1,\,p_1=1,\,p_2=\frac{3}{2},\,p_3=1,\,\xi_1^0=0,\,\xi_2^0=0,\,\xi_3^0=0$. 
  The lines are the trajectories of the arms and stem structures, and the points are the endpoints of the variable length stem structures.}
	\label{fig3-2}
\end{figure}

Comparing the proposition \ref{prop3.1}--\ref{prop3.4}, we find that the asymptotic forms of four arms in two cases have a coordinates translation, and the asymptotic forms of stem structures are the same. Therefore, the dynamical evolutions of solitons in these two cases are also very similar, as illustrated in Fig.\ \ref{fig3-2}. Next, we analyse the properties of stem structures.

\begin{table}[http]
  \centering
  \caption{Physical quantities of the arms with $k_1=k_2> k_3>0,\,0<p_1=p_3< p_2$}
  \label{tab:t3}
  \begin{tabular}{cccccc}
    \Xhline{1pt}
     Soliton & Trajectory & Velocity & Amplitude & Components \\
    \hline
    \multirow{2}{*}{$S_{1-2}$} & \multirow{2}{*}{$l_{1-2}$} & \multirow{2}{*}{$(0,\,0)$} & $0$ & $u_{1-2}$ \\
     & & &$0$ & $v_{1-2}$ \\
    \hline
    \multirow{2}{*}{$S_{1-3}$} & \multirow{2}{*}{$l_{1-3}$} & \multirow{2}{*}{$(k_1^2+k_1k_3+k_3^2,\,0)$} & $0$ & $u_{1-3}$ \\
     & & &$-\frac{(k_1-k_3)^2}{2}$ & $v_{1-3}$ \\
    \hline
    \multirow{2}{*}{$S_{1-2-3}$} & \multirow{2}{*}{$\widehat{l_{1-2-3}}$} & \multirow{2}{*}{$(k_3^2,\,-\frac{k_3^3}{p_2})$} & $\frac{k_3p_2}{2}$ & $\widehat{u_{1-2-3}}$ \\
     & & &$-\frac{k_3^2}{2}$ & $\widehat{v_{1-2-3}}$ \\
    \Xhline{1pt}
  \end{tabular}
  \caption*{\captionsetup{justification=raggedright,singlelinecheck=false,format=hang} \quad The solitons $S_j$ $(j=1-2,\,1-3,\,1-2-3)$ are composed by two components $u_{j}$ and $v_{j}$, and their trajectories are listed by \eqref{l01}.}
\end{table}

The trajectories, amplitudes, and velocities of the arms are outlined in Tables \ref{tab:t1} and \ref{tab:t3}. In this case, the amplitude of $u_{1-2}$ and $u_{1-3}$ are zero, while the amplitude of $v_{1-2}$ and $v_{1-3}$ respectively are zero and $-\frac{(k_1-k_3)^2}{2}$. And the lines are the trajectories at different moments. It can be seen from the above analysis and figures that when $t\to-\infty$, the horizontal variable length stem $S_{1-2}$ produced by the resonance gradually decreases in length until it disappears around $t=0$, and the resulting vertical variable length stem $S_{1-3}$ when $t\to+\infty$. By solving the intersection points of trajectories, the endpoints of the variable-length stem structures can be obtained as
\begin{flalign}\label{PQMN}
\begin{split}
&M\,\left(k_1^2t-\frac{p_1\xi_2^0-p_2\xi_1^0}{k_1(p_1-p_2)},\,-\frac{\xi_1^0-\xi_2^0}{p_1-p_2}\right),\,N\,\left(-k_3^2t+\frac{\ln a_{23}}{k_3}-\frac{p_1(\xi_1^0-\xi_2^0-\xi_3^0)+p_2\xi_3^0}{k_3(p_1-p_2)},\,-\frac{\xi_1^0-\xi_3^0}{p_1-p_2}\right),\\
&P\,\left((k_1^2+k_1k_3+k_3^2)t-\frac{\xi_1^0-\xi_3^0}{k_1-k_3},\,-\frac{k_1k_3(k_1+k_3)t}{p_1}+\frac{k_3\xi_1^0-k_1\xi_3^0}{p_1(k_1-k_3)}\right),\,\\
&Q\,\left((k_1^2+k_1k_3+k_3^2)t-\frac{\xi_1^0-\xi_3^0}{k_1-k_3},\,-\frac{k_1k_3(k_1+k_3)t+\ln a_{23}+\xi_2^0}{p_2}+\frac{k_1(\xi_1^0-\xi_3^0)}{p_2(k_1-k_3)}\right).
\end{split}
\end{flalign}
Here, ``M" and ``N" represent the intersections of $l_1$ and $l_2$, $\widehat{l_3}$ and $\widehat{l_{1-2-3}}$ at $t \ll 0$, while ``P" and ``Q" denote the intersections of $l_1$ and $l_3$, $\widehat{l_2}$ and $\widehat{l_{1-2-3}}$ at $t \gg 0$. These specific points are visually illustrated in Fig.\ \ref{fig3-2}. Subsequently, the lengths of the trajectories of the variable length stem structures are determined as follows:
{\small
\begin{flalign}\label{PQ+MN}
\begin{split}
&|PQ|=\left|\frac{k_1k_3(p_1-p_2)(k_1^2-k_3^2)t+p_1(k_1-k_3)\ln a_{23}-k_1p_1(\xi_1^0-\xi_2^0-\xi_3^0)-k_1p_2\xi_3^0-k_3p_1\xi_2^0+k_3p_2\xi_1^0}{p_1p_2(k_1-k_3)}\right|,\\
&|MN|=\left|\frac{k_1k_3(p_1-p_2)(k_1^2-k_3^2)t+k_1(p_1-p_2)\ln a_{23}-k_1p_1(\xi_1^0-\xi_2^0-\xi_3^0)-k_1p_2\xi_3^0-k_3p_1\xi_2^0+k_3p_2\xi_1^0}{k_1k_3(p_1-p_2)}\right|.
\end{split}
\end{flalign}}

Now, we analyze the amplitudes of variable length stem structures. 
Here we direct our attention to $v$, and the formula for $u$ corresponding to Eqs. \eqref{cross32} and \eqref{amplocal2} are provided in the appendix.  
To simplify the calculation, we choose $k_1=2,\,k_2=2,\,k_3=1,\,p_1=1,\,p_2=\frac{3}{2},\,p_3=1,\,\xi_1^0=0,\,\xi_2^0=0,\,\xi_3^0=0$. 
The cross-sectional curves of 3-soliton \eqref{uv} with Eqs.\ \eqref{3soliton1} along $l_{1-2}$ and $l_{1-3}$ shown in Fig. \ref{fig3-4} (c) and (d) are expressed as,
\begin{flalign}\label{cross32}
\begin{split}
&v|_{l_{1-2}}^{(2)}=-\frac{30(2\e^{5x-17t}+4\e^{4x-10t}+39\e^{3x-9t}+120\e^{2x-8t}+15\e^{x-t})}{(\e^{3x-9t}+30\e^{2x-8t}+15\e^{x-t}+15)^2},\\
&v|_{l_{1-3}}^{(2)}=-\frac{30(5\e^{\frac{7y}{2}+18t}+\e^{4y+18t}+15\e^{2y+12t}+24\e^{\frac{5y}{2}+12t}+75\e^{y+6t}+60\e^{\frac{3y}{2}+6t})}
{(\e^{\frac{5y}{2}+12t}+30\e^{y+6t}+15\e^{\frac{3y}{2}+6t}+15)^2}.
\end{split}
\end{flalign}

Since the extreme points and values are difficult to express explicitly, we study the amplitude at the midpoint of $MN$ and $PQ$, denoted by $R_3$ and $R_4$. 
Similar to the preceding case in this section, we exclusively showcase the amplitude trend plot for component $u$ in Fig.\ \ref{fig3-3}. 
Conversely, for component $v$, we furnish the amplitude formula and the associated trend plot over time. 
Then, the amplitudes of the variable length stem structures, respectively, are
\begin{flalign}\label{amplocal2}
\begin{split}
S_{1-2}:\quad &v(R_3)=-\frac{\alpha_6\left(2\e^{\frac{15t}{2}}+30\e^{\frac{9t}{2}}+6\alpha_6\e^{6t}+10\alpha_6\e^{3t}+75\right)}
{\left(\alpha_6\e^{\frac{9t}{2}}+\alpha_6\e^{\frac{3t}{2}}+\e^{3t}+30\right)^2},\,\\
S_{1-3}:\quad &v(R_4)=-\frac{225 \cdot \left(2\alpha_6\alpha_7\e^{\frac{5t}{2}}+ 6\alpha_7 \e^{4t} + 9\alpha_8 \e^{\frac{7t}{2}}+ 6\alpha_6 \e^{\frac{3t}{2}} + 2 \alpha_7\e^t\right)}
{\left(\alpha_6\alpha_7\e^t+30\alpha_7\e^{\frac{5t}{2}} + 15 \e^{\frac{3t}{2}} + 15\sqrt{15} \right)^2},
\end{split}
\end{flalign}
where $\alpha_6=\sqrt{15},\,\alpha_7=\sqrt[3]{15},\,\alpha_8=\sqrt[6]{15}$.

The limits are easily obtained:
$$\lim_{t\to +\infty} v(R_3)=0,\,\lim_{t\to -\infty} v(R_3)=0,\,\lim_{t\to +\infty} v(R_4)=-\frac{1}{2},\,\lim_{t\to -\infty} v(R_4)=0.$$
These results align with the observations in Fig.\ \ref{fig3-3}. The figures of \eqref{amplocal2} are shown in Fig.\ \ref{fig3-3} (b). As can be seen from the figure that $S_{1-2}$ disappears around $t=0$, and $S_{1-3}$ arises.

\section{The stem structure in 3-soliton generated by strong 2-resonance in soliton reconnection}\label{sec4}
In this section, we shall use two-variable asymptotic method to study the 3-soliton with strong 2-resonance. 
By substituting $\xi_2^0\to\xi_2^0-\ln a_{12}$ and $\xi_3^0\to\xi_3^0-\ln a_{13}$ (equivalent to $\xi_2\to\xi_2-\ln a_{12}$ and $\xi_3\to\xi_3-\ln a_{13}$) into Eq.\ \eqref{3soliton} and taking the limit as $a_{12},\,a_{13}\to \infty$, we can derive
\begin{flalign}\label{3soliton2}
\begin{split}
	f_{strong}=1+\exp\xi_1+\exp(\xi_1+\xi_2)+\exp(\xi_1+\xi_3)+a_{23}\exp(\xi_1+\xi_2+\xi_3).
\end{split}
\end{flalign}
The 3-soliton with strong 2-resonance is precisely described by Eqs.\ \eqref{3soliton2} and \eqref{uv}. To ensure that $a_{12}=\infty$, $a_{13}=\infty$, and $0<a_{23}<+\infty$, two distinct cases emerge: (i) $0<k_1=-k_3<k_2,\,p_1=-p_2>p_3>0$; (ii) $k_1=-k_2>k_3>0,\,0<p_1=-p_3<p_2$. Next, we will analyze these two cases separately.
\begin{table}[http]
  \centering
   \caption{Physical quantities of the arms with $0<k_1=-k_3<k_2,\,p_1=-p_2>p_3>0$}\label{tab:t4}
  \begin{tabular}{cccccc}
\Xhline{1pt}
    Soliton &Trajectory & Velocity  &Amplitude &Components\\
     \hline
    \multirow{2}{*}{$S_{1+2}$} &\multirow{2}{*}{$l_{1+2}$} &\multirow{2}{*}{$(k_1^2-k_1k_2+k_2^2,\,0)$}  &$0$ &$u_{1+2}$\\
      & &  &$-\frac{(k_1+k_2)^2}{2}$ &$v_{1+2}$\\
    \hline
    \multirow{2}{*}{$S_{1+3}$} &\multirow{2}{*}{$l_{1+3}$} &\multirow{2}{*}{$(0,\,0)$} &$0$ &$u_{1+3}$ \\
     & &  &$0$ &$v_{1+3}$ \\
    \hline
    \multirow{2}{*}{$S_{1+2+3}$} &\multirow{2}{*}{$\widehat{l_{1+2+3}}$} &\multirow{2}{*}{$(k_2^2,\,-\frac{k_2^3}{p_3})$}  &$\frac{k_2p_3}{2}$ &$\widehat{u_{1+2+3}}$ \\
     & & &$-\frac{k_2^2}{2}$  &$\widehat{v_{1+2+3}}$\\
   \Xhline{1pt}
  \end{tabular}
  \caption*{\captionsetup{justification=raggedright,singlelinecheck=false,format=hang} \quad The solitons $S_j$ $(j=1+2,\,1+3,\,1+2+3)$ are composed by two components $u_{j}$ and $v_{j}$, and their trajectories are listed by \eqref{l02}.}
\end{table}

\subsection{The asymptotic analysis with $0<k_1=-k_3<k_2,\,p_1=-p_2>p_3>0$}
First we use a similar method to the previous section for asymptotic analysis.

{\bf I.} On the region $\eta_1=x+\frac{p_1}{k_1}y-k_1^2t$, we have
\begin{align*}
&\xi_1=k_1\eta_1+c_1,\,\xi_1+\xi_2=(k_1+k_2)\eta_1+\frac{k_1p_2-k_2p_1}{k_1}y+k_2(k_1^2-k_2^2)t+c_2,\\
&\xi_1+\xi_3=(k_1+k_3)\eta_1+\frac{k_1p_3-k_3p_1}{k_1}y+c_3,\\
&\xi_1+\xi_2+\xi_3=(k_1+k_2+k_3)\eta_1+\frac{(p_2 + p_3)k_1 - p_1(k_2 + k_3)}{k_1}y+((k_2+k_3)k_1^2-(k_2^3+k_3^3))t+c_4.
\end{align*}

{\bf (a)} In the case of $t\to +\infty$:  We have $\xi_1\approx c,\,\xi_1+\xi_2\to -\infty,\,\xi_1+\xi_3\approx c,\,\xi_1+\xi_2+\xi_3\to -\infty$, and then $f\sim 1+\e^{\xi_1}+\e^{\xi_1+\xi_3}$. Further we can get:
$$f\sim 1+\e^{\xi_1},\,y\to -\infty;\,f\sim \e^{\xi_1}+\e^{\xi_1+\xi_3},\,y\to +\infty;\,f\sim 1+\e^{\xi_1+\xi_3},\,x\to -\infty.$$

{\bf (b)} In the case of $t\to -\infty$:  We have $\xi_1\approx c,\,\xi_1+\xi_2\to +\infty,\,\xi_1+\xi_3\to -\infty,\,\xi_1+\xi_2+\xi_3\to +\infty$, and then $f\sim \e^{\xi_1+\xi_2}+a_{23}\e^{\xi_1+\xi_2+\xi_3}$. It can only be determined if the arm appears at $t\to -\infty$, not whether it is on $y\to +\infty$ or $y\to -\infty$.

{\bf II.} On the region $\eta_2=x+\frac{p_2}{k_2}y-k_2^2t$, we have
\begin{align*}
&\xi_1=k_1\eta_2+\frac{k_2p_1-k_1p_2}{k_2}y+k_1(k_2^2-k_1^2)t+c_1,\\
&\xi_1+\xi_2=(k_1+k_2)\eta_2+\frac{k_2p_1-k_1p_2}{k_2}y+k_1(k_2^2-k_1^2)t+c_2,\\
&\xi_1+\xi_3=(k_1+k_3)\eta_2+(p_1+p_3)y+k_2^2(k_1+k_3)t+c_3,\\
&\xi_1+\xi_2+\xi_3=(k_1+k_2+k_3)\eta_2+(p_1+p_3)y+k_2^2(k_1+k_3)t+c_4.
\end{align*}

{\bf (a)} In the case of $t\to -\infty$:  We have $\xi_2\approx c,\,\xi_1\to -\infty,\,\xi_1+\xi_2\to -\infty,\,\xi_1+\xi_3\approx c,\,\xi_1+\xi_2+\xi_3\approx c$, and then $f\sim 1+\e^{\xi_1+\xi_3}+a_{23}\e^{\xi_1+\xi_2+\xi_3}$. Further we can get:
$$f\sim \e^{\xi_1+\xi_3}+a_{23}\e^{\xi_1+\xi_2+\xi_3},\,y\to +\infty;\,f\sim 1+a_{23}\e^{\xi_1+\xi_2+\xi_3},\,y\to-\infty;\,f\sim 1+\e^{\xi_1+\xi_3},\,x\to-\infty.$$

{\bf (b)} In the case of $t\to +\infty,\,y\to +\infty$: 
By the limit $y\to +\infty$, we an get $\xi_2\approx c,\,\xi_1\to +\infty,\,\xi_1+\xi_2\to +\infty,\,\xi_1+\xi_3\to +\infty,\,\xi_1+\xi_2+\xi_3\to +\infty$. So $f\sim \e^{\xi_1}+\e^{\xi_1+\xi_2}+\e^{\xi_1+\xi_3}+a_{23}\e^{\xi_1+\xi_2+\xi_3}=\e^{\xi_1}(1+\e^{\xi_2}+\e^{\xi_3}+a_{23}\e^{\xi_2+\xi_3})$. Due to $\xi_3\to -\infty$, it can be get as following
$$f\sim 1+\e^{\xi_2},\, t\to +\infty,\,y\to +\infty.$$

{\bf III.} On the region $\eta_3=x+\frac{p_3}{k_3}y-k_3^2t$, we have
\begin{align*}
&\xi_1=k_1\eta_3+\frac{k_3p_1-k_1p_3}{k_3}y+c_1,\\
&\xi_1+\xi_2=(k_1 + k_2)\eta_3+\frac{k_2(p_1+p_3)}{k_3}y+k_2(k_3^2-k_2^2)t+c_2,\\
&\xi_1+\xi_3=(k_1 + k_3)\eta_3+\frac{k_3p_1-k_1p_3}{k_3}y+c_3,\\
&\xi_1+\xi_2+\xi_3=(k_1 +k_2+k_3)\eta_1+\frac{k_2(p_1+p_3)}{k_3}y+k_2(k_3^2-k_2^2)t+c_4.
\end{align*}

{\bf (a)} In the case of $t\to +\infty$:  We have $\xi_3\approx c,\,\xi_1\approx c,\,\xi_1+\xi_2\to -\infty,\,\xi_1+\xi_3\approx c,\,\xi_1+\xi_2+\xi_3\to -\infty$, and then $f\sim 1+\e^{\xi_1}+\e^{\xi_1+\xi_3}$. This situation is the same as the case {\bf I (a)} above.

{\bf (b)} In the case of $t\to -\infty,\,y\to +\infty$: We have $\xi_3\approx c,\,\xi_1\to +\infty,\,\xi_1+\xi_2\to +\infty,\,\xi_1+\xi_3\to +\infty,\,\xi_1+\xi_2+\xi_3\to +\infty$. Because of $\xi_2\gg\xi_3$ when $t\to -\infty$, so we have
$$f\sim \e^{\xi_1+\xi_2}+a_{23}\e^{\xi_1+\xi_2+\xi_3},\,t\to -\infty,\,y\to +\infty.$$

{\bf IV.} On the region $\eta_4=x+\frac{p_1+p_2}{k_1+k_2}y-(k_1^3+k_2^2)t$, we have
\begin{align*}
&\xi_1=k_1\eta_3+\frac{k_3p_1-k_1p_3}{k_3}y+c_1,\\
&\xi_1+\xi_2=(k_1 + k_2)\eta_3+\frac{k_2(p_1+p_3)}{k_3}y+k_2(k_3^2-k_2^2)t+c_2,\\
&\xi_1+\xi_3=(k_1 + k_3)\eta_3+\frac{k_3p_1-k_1p_3}{k_3}y+c_3,\\
&\xi_1+\xi_2+\xi_3=(k_1 +k_2+k_3)\eta_1+\frac{k_2(p_1+p_3)}{k_3}y+k_2(k_3^2-k_2^2)t+c_4.
\end{align*}

{\bf (a)} In the case of $y\to -\infty$:  We have $\xi_1\to -\infty,\,\xi_1+\xi_2\approx c,\,\xi_1+\xi_3\to -\infty,\,\xi_1+\xi_2+\xi_3\to -\infty$, and then $f\sim 1+\e^{\xi_1+\xi_2}$. Considering the analysis of cases \textbf{I}--\textbf{III} presented in this section, and noting that soliton $S_{1+2}$ arises from a strong resonance between $S_1$ and $S_2$ ($a_{12}\to+\infty$), it is evident that there is no phase shift as $t\to-\infty$ or $t\to+\infty$. Hence, it can be inferred that this soliton persists indefinitely as $t\to\pm\infty$, namely,
$$f\sim 1+\e^{\xi_1+\xi_2},\,y\to -\infty,\,t\to\pm\infty.$$

We notice that some of the asymptotic forms obtained from the above analysis are repeated. 
After sorting, the asymptotic forms are as follows:

Before collision ($t\to -\infty$):
\begin{flalign}
\begin{split}
&f_2^-\sim 1+a_{23}\e^{\eta_2},\,f_3^-\sim 1+a_{23}\e^{\eta_3},\,(y\to +\infty),\\
&f_{1+2}\sim 1+\e^{\eta_1+\eta_2},\,(y\to -\infty),\\
&f_{1+3}\sim 1+\e^{\eta_1+\eta_3},\,(x\to -\infty).
\end{split}
\end{flalign}

After collision ($t\to +\infty$):
\begin{flalign}
\begin{split}
&f_2^+\sim 1+\e^{\eta_2},\,f_3^+\sim 1+\e^{\eta_3},\,(y\to +\infty),\\
&f_{1+2}\sim 1+\e^{\eta_1+\eta_2},\,(y\to -\infty),\\
&f_{1+3}\sim 1+\e^{\eta_1+\eta_3},\,(x\to -\infty).
\end{split}
\end{flalign}

Then we have the following proposition:
\begin{prop}\label{prop4.1}
The asymptotic forms of the strong 2-resonant 3-soliton with $0<k_1=-k_3<k_2,\,p_1=-p_2>p_3>0$ are as following:

Before collision ($t\to-\infty$):
\begin{flalign}\label{asy05}
\begin{split}
x\to +\infty,\,&S_2:\quad \widehat{u_2}\approx -\frac{k_2p_2}{2}\sech^2(\frac{\xi_2+\ln a_{23}}{2}),\,\widehat{v_2}\approx -\frac{k_2^2}{2}\sech^2(\frac{\xi_2}{2}),\\
&S_3:\quad \widehat{u_3}\approx -\frac{k_3p_3}{2}\sech^2(\frac{\xi_3+\ln a_{23}}{2}),\,\widehat{v_3}\approx -\frac{k_3^2}{2}\sech^2(\frac{\xi_3+ \ln a_{23}}{2}),\\
x\to -\infty,\,&S_{1+3}:\, u_{1+3}\approx -\frac{(k_1+k_3)(p_1+p_3)}{2}\sech^2(\frac{\xi_1+\xi_3}{2}),\,v_{1+3}\approx -\frac{(k_1+k_3)^2}{2}\sech^2(\frac{\xi_1+\xi_3}{2}),\\
y\to -\infty,\,&S_{1+2}:\, u_{1+2}\approx -\frac{(k_1+k_2)(p_1+p_2)}{2}\sech^2(\frac{\xi_1+\xi_2}{2}),\,v_{1+2}\approx -\frac{(k_1+k_2)^2}{2}\sech^2(\frac{\xi_1+\xi_2}{2}).
\end{split}
\end{flalign}

Before collision ($t\to+\infty$):
\begin{flalign}\label{asy06}
\begin{split}
x\to +\infty,\,&S_2:\quad u_2\approx -\frac{k_2p_2}{2}\sech^2(\frac{\xi_2}{2}),\,v_2\approx -\frac{k_2^2}{2}\sech^2(\frac{\xi_2}{2}),\\
&S_3:\quad u_3\approx -\frac{k_3p_3}{2}\sech^2(\frac{\xi_3}{2}),\,v_3\approx -\frac{k_3^2}{2}\sech^2(\frac{\xi_3}{2}).\\
x\to -\infty,\,&S_{1+3}:\, u_{1+3}\approx -\frac{(k_1+k_3)(p_1+p_3)}{2}\sech^2(\frac{\xi_1+\xi_3}{2}),\,v_{1+3}\approx -\frac{(k_1+k_3)^2}{2}\sech^2(\frac{\xi_1+\xi_3}{2}),\\
y\to -\infty,\,&S_{1+2}:\, u_{1+2}\approx -\frac{(k_1+k_2)(p_1+p_2)}{2}\sech^2(\frac{\xi_1+\xi_2}{2}),\,v_{1+2}\approx -\frac{(k_1+k_2)^2}{2}\sech^2(\frac{\xi_1+\xi_2}{2}).
\end{split}
\end{flalign}
\end{prop}

\begin{rk}
It can be seen from \eqref{asy05} and \eqref{asy06} that the asymptotic forms of the four arms are partially changed with respect to $t$: The asymptotic form of $S_2$ and $S_3$ differ by a phase shift $\ln a_{23}$ at $t\to-\infty$ and $t \to +\infty$. If $a_{23}=1$, the asymptotic forms \eqref{asy05} and \eqref{asy06} are the same, and the four arms shift over time without a phase shift. The same is true of the asymptotic forms \eqref{asy07} and \eqref{asy08} in section 4.3.
\end{rk}

And then we consider the two special arms $S_1$ and $S_{1+2+3}$ which respectively associated with $f\sim 1+\e^{\xi_1}\,(t\to+\infty)$ and $f\sim 1+a_{23}\e^{\xi_1+\xi_2+\xi_3}\,(t\to-\infty)$ in the asymptotic analysis above. When $t\to +\infty$, the resonance between $S_1$ and $S_2$ leads to the $S_{1+2}$, the resonance between $S_1$ and $S_3$ leads to the $S_{1+3}$. When $t\to -\infty$, the resonances between $S_2$ and $S_{1+3}$ leads to the $S_{1+2+3}$ which can also be viewed as generated by the resonance between $S_3$ and $S_{1+2}$. Considering the asymptotic trend of the four arms, we can see that $S_1$ and $S_{1+2+3}$ actually local structures of finite length which named by stem structures in this paper. Then we have asymptotic forms of the stems as,
\begin{flalign*}
f_1 \sim 1+\e^{\xi_1},\,t\to +\infty,\,\text{and}\,f_{1+2+3}\sim 1+a_{23}\e^{\xi_1+\xi_2+\xi_3},\,t\to -\infty.
\end{flalign*}
Namely,
\begin{prop}\label{prop4.2}
The stem structures corresponding to asymptotic forms \eqref{asy05} and \eqref{asy06} are as following:
\begin{flalign}
\begin{split}
t\to+\infty,\,S_1:\quad\,\, &u_1\approx -\frac{k_1p_1}{2}\sech^2(\frac{\xi_1}{2}),\,v_1\approx -\frac{k_1^2}{2}\sech^2(\frac{\xi_1}{2}),\\
t\to-\infty,\,S_{1+2+3}:\,&\widehat{u_{1+2+3}}\approx -\frac{(k_1+k_2+k_3)(p_1+p_2+p_3)}{2}\sech^2(\frac{\xi_1+\xi_2+\xi_3+\ln a_{23}}{2}),\\
&\widehat{v_{1+2+3}}\approx -\frac{(k_1+k_2+k_3)^2}{2}\sech^2(\frac{\xi_1+\xi_2+\xi_3}{2}).
\end{split}\label{stem03}
\end{flalign}
\end{prop}

\begin{figure}[h!tb]
	\centering
    \subfigure[$u:\,t=-8$]{\includegraphics[height=3.4cm,width=3.4cm]{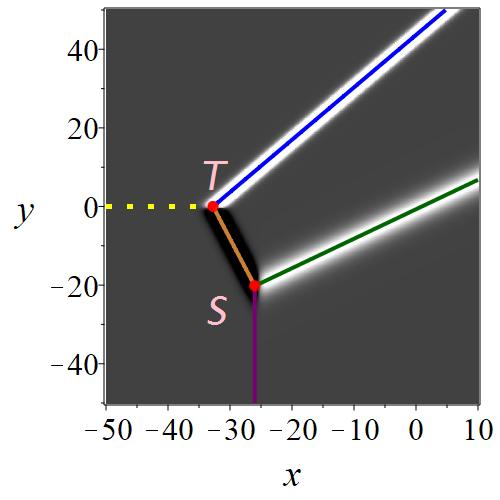}}
    \subfigure[$u:\,t=-4$]{\includegraphics[height=3.4cm,width=3.4cm]{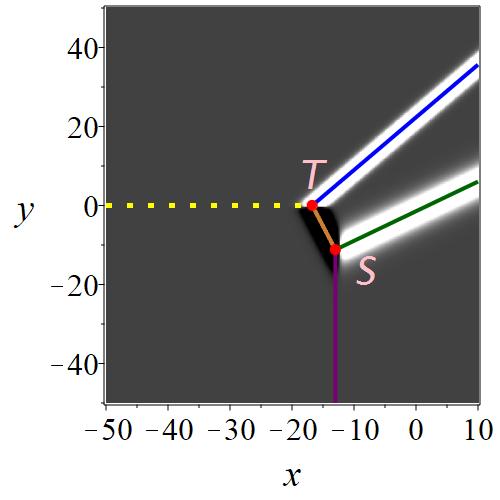}}
	\subfigure[$u:\,t=0$]{\includegraphics[height=3.4cm,width=3.4cm]{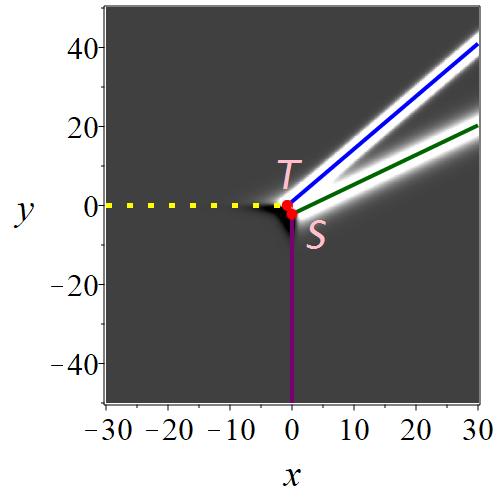}}
	\subfigure[$u:\,t=5$]{\includegraphics[height=3.4cm,width=3.4cm]{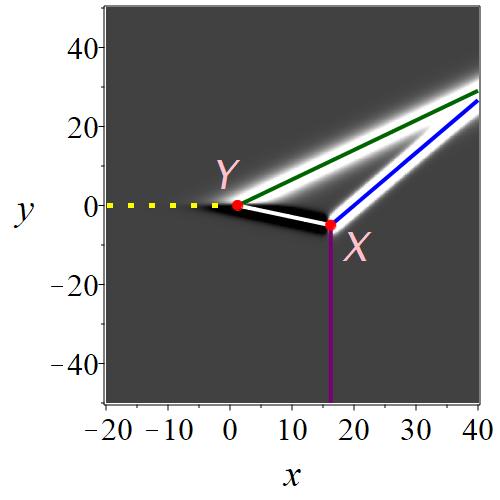}}
    \subfigure[$u:\,t=10$]{\includegraphics[height=3.4cm,width=3.4cm]{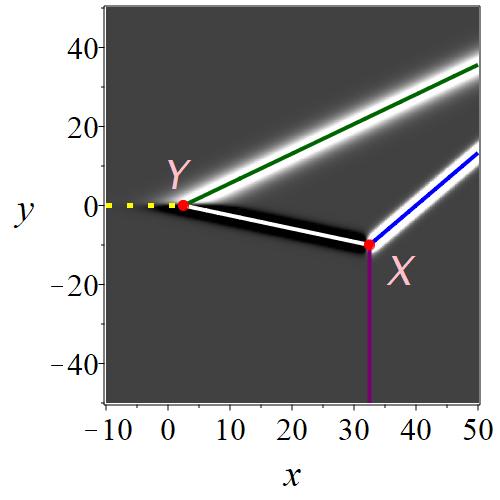}}\\
    \subfigure[$v:\,t=-8$]{\includegraphics[height=3.4cm,width=3.4cm]{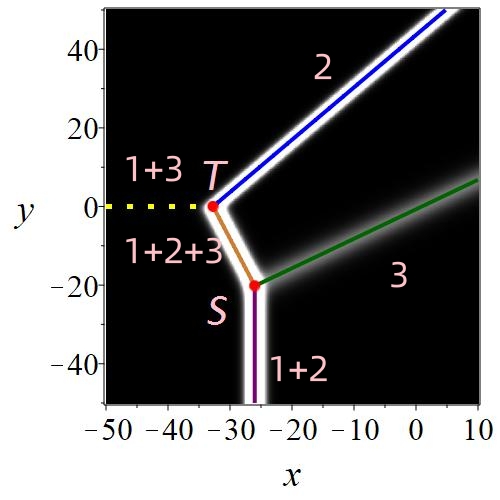}}
    \subfigure[$v:\,t=-4$]{\includegraphics[height=3.4cm,width=3.4cm]{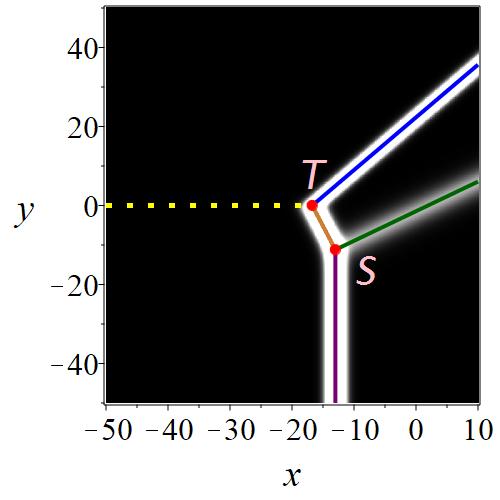}}
	\subfigure[$v:\,t=0$]{\includegraphics[height=3.4cm,width=3.4cm]{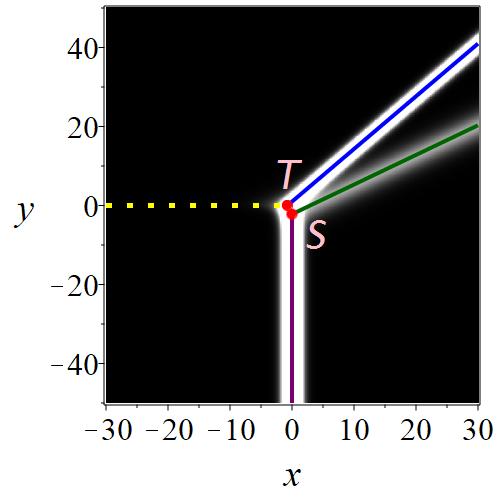}}
	\subfigure[$v:\,t=5$]{\includegraphics[height=3.4cm,width=3.4cm]{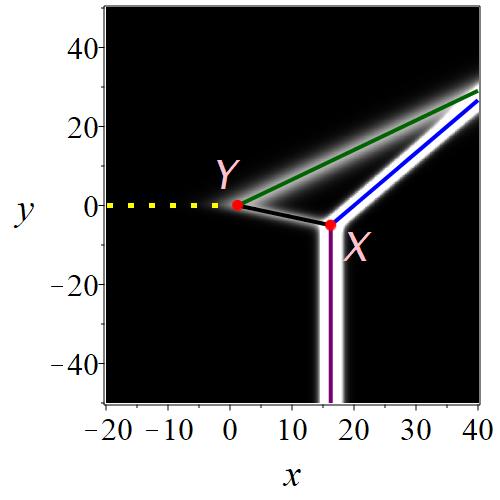}}
    \subfigure[$v:\,t=10$]{\includegraphics[height=3.4cm,width=3.4cm]{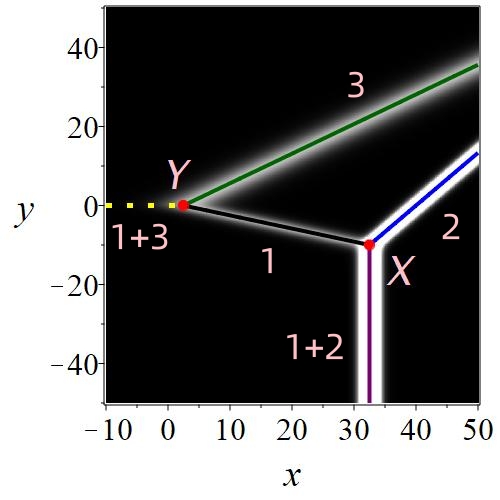}}
	\caption{The density plots of the strong 2-resonant 3-soliton given by \eqref{uv} and \eqref{3soliton2} with  $k_1=\frac{1}{2},\,k_2=2,\,k_3=-\frac{1}{2},\,p_1=\frac{3}{2},\,p_2=-\frac{3}{2},\,p_3=\frac{2}{3},\,\xi_1^0=0,\,\xi_2^0=0,\,\xi_3^0=0$.  The lines are the trajectories of the arms and stem structures, and the points are the endpoints of the variable length stem structures.
	}\label{fig3-5}
\end{figure}

According to proposition \ref{prop4.1} and \ref{prop4.2}, the asymptotic forms indicate that the 2-resonant 3-soliton features a configuration with four arms and a central stem. This temporal evolution is depicted in Fig.\ \ref{fig3-5}. As $t \to -\infty$, the stem structure $S_{1+2+3}$ links two pairs of V-shaped solitons: $S_2$ and $S_{1+3}$, $S_3$ and $S_{1+2}$. Over time, this stem structure shortens and eventually disappears around $t=0$. At this critical moment, the four arms ($S_2,\,S_3,\,S_{1+2},\,S_{1+3}$) converge, transforming the pairs of V-shaped solitons into $S_2$ and $S_{1+2}$, $S_3$ and $S_{1+3}$. As time proceeds ($t \to +\infty$), a new stem structure $S_1$ forms and elongates, reconnecting these pairs of V-shaped solitons. This is the full process of the soliton reconnection induced by
the 2-resonance.

\subsection{The stem structures with $0<k_1=-k_3<k_2,\,p_1=-p_2>p_3>0$}
The trajectory, amplitude, and velocity of the four arms and two stems are outlined in Tables \ref{tab:t1} and \ref{tab:t4}, where the trajectories are given by \eqref{l01} and below formulas,
\begin{flalign}\label{l02}
\begin{split}
\boldsymbol{l_{1+2}:}\, \xi_1+\xi_2=0,\,\boldsymbol{l_{1+3}:}\,\xi_1+\xi_3=0;\,\boldsymbol{\widehat{l_{1+2+3}}:}\,\xi_1+\xi_2+\xi_3+\ln a_{23}=0.
\end{split}
\end{flalign}

It's easy to demonstrate that the trajectories of $S_{1+2}$, $S_3$, and $S_{1+2+3}$ intersect at a single point, and that the trajectories of $S_2$, $S_{1+3}$, and $S_{1+2+3}$ intersect at a single point as $t \to -\infty$. Similarly, the trajectories of $S_1$, $S_3$, and $S_{1+3}$ intersect at a single point, and the trajectories of $S_1$, $S_2$, and $S_{1+2}$ intersect at a single point as $t \to +\infty$. This further confirms the accuracy of the stem structure positions given in \eqref{stem03}. It is noted that $S_{1+3}$ has a vertical orientation, while $S_{1+2}$ has a horizontal alignment. The trajectories of the arms and stem structures are illustrated in Fig.\ \ref{fig3-5}. By solving for the intersection points of these trajectories, the endpoints of the variable-length stem structures can be determined as follows:
\begin{flalign}\label{STXY}
	\begin{split}
&S\,\left(\frac{(p_1+p_2)(h_3+\ln a_{23})-p_3(h_1+h_2)}{p_3(k_1+k_2)-k_3(p_1+p_2)},\,\frac{-(k_1+k_2)(h_3+\ln a_{23})+k_3(h_1+h_2)}{p_3(k_1+k_2)-k_3(p_1+p_2)}\right),\\
&T\,\left(\frac{(p_1+p_3)(h_2+\ln a_{23})-p_2(h_1+h_3)}{p_2(k_1+k_3)-k_2(p_1+p_3)},\,\frac{-(k_1+k_3)(h_2+\ln a_{23})+k_2(h_1+h_3)}{p_2(k_1+k_3)-k_2(p_1+p_3)}\right),\\
&X\,\left(\frac{p_1h_2-p_2h_1}{p_2k_1-p_1k_2},\,\frac{-k_1h_2+k_2h_1}{p_2k_1-p_1k_2}\right),\quad
Y\,\left(\frac{p_1h_3-p_3h_1}{p_3k_1-p_1k_3},\,\frac{-k_1h_3+k_3h_1}{p_3k_1-p_1k_3}\right),
	\end{split}
\end{flalign}
where $h_1=-k_1^3t+\xi_1^0,\,h_2=-k_2^3t+\xi_2^0,\,h_3=-k_3^3t+\xi_3^0$. ``S" and ``T" respectively are the intersections of $l_{1+2},\,\widehat{l_3}$ and $\widehat{l_2},\,\widehat{l_{1+2+3}}$ at $t\ll 0$, while ``X" and ``Y" respectively are the intersections of $l_{1+2},\,l_2$ and $l_1,\,l_3$ at $t\gg 0$. Whereupon, the lengths of the variable length stem structures are obtained as following,
\begin{flalign}\label{ST+XY}
\begin{split}
&|ST|=\left|\frac{(k_2p_3-k_3p_2)(h_1-\ln a_{23})+(k_3p_1-k_1p_3)(h_2+\ln a_{23})+(k_1p_2-k_2p_1)(h_3+\ln a_{23})}
{(p_3(k_1+k_2)-k_3(p_1+p_2))(p_2(k_1+k_3)-k_2(p_1+p_3))}\right|\\
&\quad\cdot\sqrt{(k_1+k_2+k_3)^2+(p_1+p_2+p_3)^2},\\
&|XY|=\left|\frac{(k_2p_3-k_3p_2)h_1+(k_3p_1-k_1p_3)h_2+(k_1p_2-k_2p_1)h_3}{(k_1p_2 - k_2p_1)(k_1p_3 - k_3p_1)}\right|\cdot\sqrt{k_1^2+p_1^2}.
\end{split}
\end{flalign}
\begin{rk}
Due to the intricate nature of the evolution around $t=0$, the expressions \eqref{STXY} and \eqref{ST+XY} are applicable exclusively in the regimes where $t\ll 0$ and $t\gg 0$.
\end{rk}

\begin{figure}[h!tb]
	\centering
	\subfigure[$v|_{l_{1+2+3}}^{(1)}:\,t=-8$]{\includegraphics[height=3.4cm,width=3.4cm]{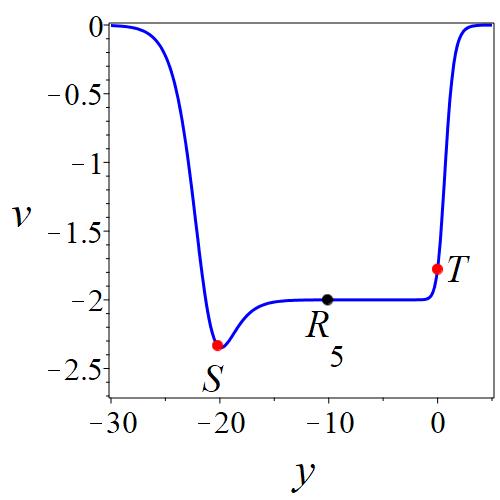}}
    \subfigure[$v|_{l_1}^{(1)}:\,t=5$]{\includegraphics[height=3.4cm,width=3.4cm]{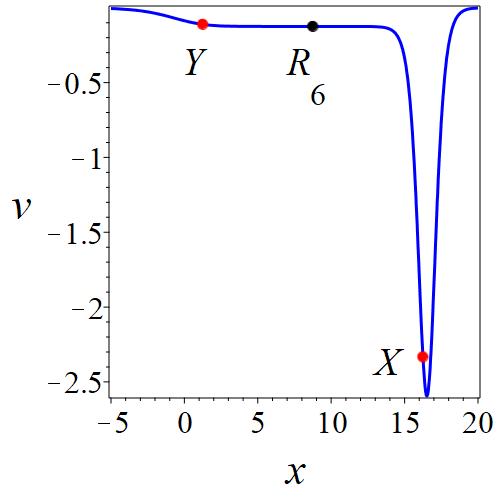}}
    \subfigure[$v|_{l_1}^{(2)}:\,t=-8$]{\includegraphics[height=3.4cm,width=3.4cm]{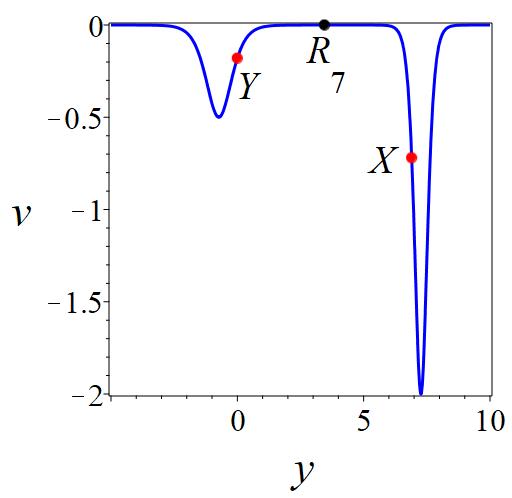}}
    \subfigure[$v|_{l_{1+2+3}}^{(2)}:\,t=5$]{\includegraphics[height=3.4cm,width=3.4cm]{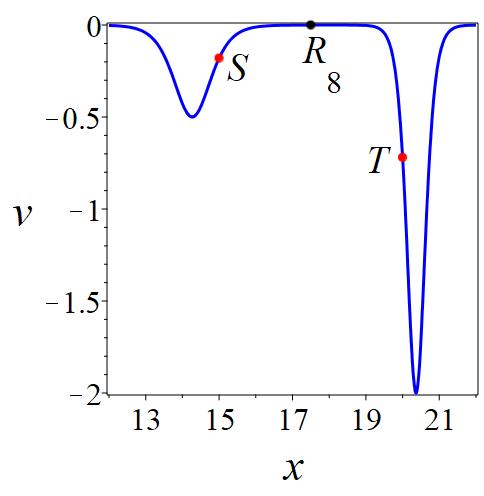}}
	\caption{(a) (b) The cross-sectional curves \eqref{cross33};
   (c) (d) The cross-sectional curves \eqref{cross34}.
   The red points are the endpoints of the variable length stem structures.}
	\label{fig3-7}
\end{figure}

Then we consider how the amplitudes of the variable length stem structures change over time. 
As the correlation analysis approaches for u and $v$ are largely the same, only the analysis of $v$ is conducted herein. 
Subsequent sections present solely the formulas associated with $v$, while those pertaining to $u$ are available in the appendix. 
For ease of calculation, we set $k_1=\frac{1}{2},\,k_2=2,\,k_3=-\frac{1}{2},\,p_1=\frac{3}{2},\,p_2=-\frac{3}{2},\,p_3=\frac{2}{3},\,\xi_1^0=0,\,\xi_2^0=0,\,\xi_3^0=0$. 
The cross-sectional curves of 3-soliton \eqref{uv} with Eqs.\ \eqref{3soliton2} along $\widehat{l_{1+2+3}}$ shown in Fig. \ref{fig3-7} (a) and (b) and $l_{1}$ are expressed as,
\begin{flalign}\label{cross33}
\begin{split}
&v|_{\widehat{l_{1+2+3}}}^{(1)}=-\frac{169 \left( 48 \sqrt{39} \e^{\frac{15t}{4} + \frac{y}{2}} + 13\beta_1\e^{\frac{15t}{8} + \frac{7y}{2}} + 
    205\beta_1\e^{\frac{15t}{8} + \frac{4y}{3}} + 78\beta_1\e^{\frac{15t}{8} - \frac{5y}{6}} + 2704 \e^{\frac{13y}{6}} + 2704 \right)}{2 \left( 
    13 \beta_1\e^{\frac{15t}{8} + \frac{4y}{3}} + 3 \beta_1\e^{\frac{15t}{8} - \frac{5y}{6}} + 169 \e^{\frac{13y}{6}} + 338 \right)^2},\\
&v|_{l_{1}}^{(1)}=-\frac{1200\e^{-\frac {143t}{18}+\frac{16x}{9}}+39\e^{-\frac{1157t}{72}+\frac{77x}{18}}+624\e^{-{\frac {559t}{72}}
+{\frac {19x}{18}}}+369\,{{\rm e}^{-{\frac {65t}{8}}+\frac{5x}{2}}}+9{\e^{{\frac {13\,t}{72}}-{\frac {13\,x}{18}}}}+9}
{2\left(3\,{\e^{-{\frac {65\,t}{8}}+\frac{5x}{2}}}+3{\e^{{\frac {13\,t}{72}}-{\frac{13x}{18}}}}+13\e^{-\frac {143t}{18}+\frac{16x}{9}}+6\right)^2},
\end{split}
\end{flalign}
where, $\beta_1=13^{\frac{3}{4}} \cdot 3^{\frac{1}{4}}$.

\begin{figure}[h!tb]
	\centering
    \raisebox{15ex}{(a)}\includegraphics[height=3.8cm,width=3.5cm]{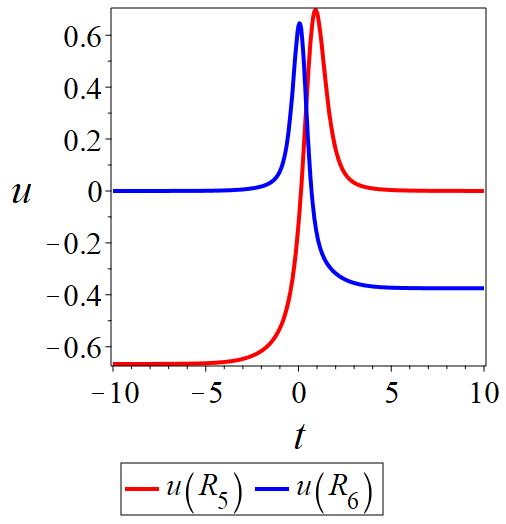}
    \raisebox{15ex}{(b)}\includegraphics[height=3.8cm,width=3.5cm]{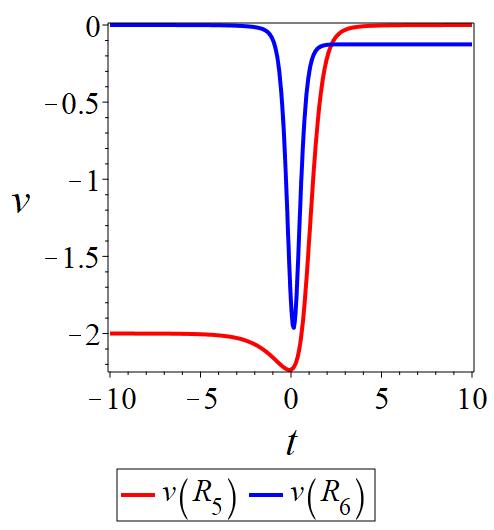}
    \raisebox{15ex}{(c)}\includegraphics[height=3.8cm,width=3.5cm]{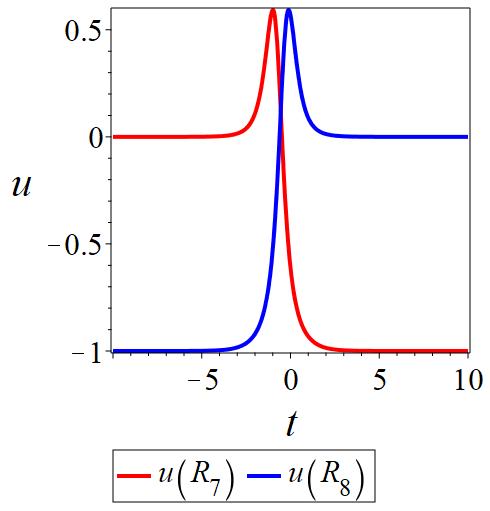}
    \raisebox{15ex}{(d)}\includegraphics[height=3.8cm,width=3.5cm]{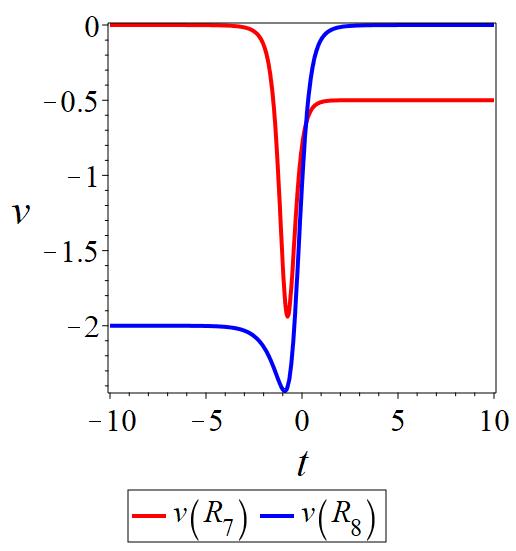}
	\caption{The evolutions of the amputation on the point $R_j$ over time: 
  (a)--(d) correspond to \eqref{amplocaladd3} \eqref{amplocal3} \eqref{amplocaladd4}\eqref{amplocal4}, respectively.
	}\label{fig3-8}
\end{figure}

The amplitudes of $u(R_5)$ and $u(R_6)$ are not given explicitly for the same reasons as the previous section. 
We just show the amplitudes of $S_{1+2+3}$ and $S_{1}$ for $u$ in Fig.\ \ref{fig3-8} and mainly discuss $v$. 
The amplitudes of $R_5$ and $R_6$ for $v$ are given as following,
\begin{flalign}\label{amplocal3}
\begin{split}
S_{1+2+3}:\, &v(R_5)=-\frac{13 \left( 9 \beta_2\e^{\frac{93t}{16}} + 624 \beta_2 \e^{\frac{69t}{16}} + 615\beta_1\e^{\frac{27t}{8}} + 624\beta_3\e^{\frac{39t}{16}} + 
4394 \beta_3\e^{\frac{15t}{16}} + 35152 \right)}
{2 \left( 3\beta_1\e^{\frac{27t}{8}} + 3 \beta_3\e^{\frac{39t}{16}} + 13\beta_3\e^{\frac{15t}{16}} + 338 \right)^2},\\
S_1:\quad &v(R_{6})=-{ \frac{3\left( 13\,{{\rm e}^{-{\frac {103\,t}{12}}}}+208\,{
{\rm e}^{-{\frac {71\,t}{12}}}}+400\,{{\rm e}^{-{\frac {29\,t}{6}}}}+
123\,{\e^{-{\frac {15\,t}{4}}}}+3\,{{\rm e}^{-{\frac {13\,t}{12}}
}}+3 \right)}{ 2 \left( 6+3\,{{\rm e}^{-{\frac {15\,t}{4}}}}+3\,{{\rm e}^
{-{\frac {13\,t}{12}}}}+13\,{{\rm e}^{-{\frac {29\,t}{6}}}} \right) ^{2}}},
\end{split}
\end{flalign}
where, $\beta_1=13^{\frac{3}{4}} \cdot 3^{\frac{1}{4}},\,\beta_2=13^{\frac{1}{8}} \cdot 3^{\frac{7}{8}},\,\beta_3=13^{\frac{3}{8}} \cdot 3^{\frac{5}{8}}$.

It is straight forward to determine the limits:
$$\lim_{t\to +\infty} v(R_5)=0,\,\lim_{t\to -\infty} v(R_5)=-2,\,\lim_{t\to +\infty} v(R_6)=-\frac{1}{8},\,\lim_{t\to -\infty} v(R_6)=0.$$
These results are exhibited by Fig.\ \ref{fig3-8}. 
As can be seen from the figure that the evolution of arms around $t=0$ is complicated. 
The figure also confirms that $S_{1+2+3}$ disappears around $t=0$ ($v(R_5)\approx 0$), and $S_1$ arises ($v(R_6)\approx 0$).
\\
\begin{table}[http]
  \centering
   \caption{Physical quantities of the arms with $k_1=-k_2>k_3>0,\,0<p_1=-p_3<p_2$}\label{tab:t5}
  \begin{tabular}{cccccc}
\Xhline{1pt}
    Soliton &Trajectory & Velocity  &Amplitude &Components \\
    \hline
    \multirow{2}{*}{$S_{1+2}$} &\multirow{2}{*}{$l_{1+2}$} &\multirow{2}{*}{$(0,\,0)$} &$0$ &$u_{1+2}$ \\
       & &  &$0$ &$v_{1+2}$\\
    \hline
    \multirow{2}{*}{$S_{1+3}$} &\multirow{2}{*}{$l_{1+3}$} &\multirow{2}{*}{$(k_1^2-k_1k_3+k_3^2,\,0)$}  &$0$ &$u_{1+3}$\\
      & &  &$-\frac{(k_1+k_3)^2}{2}$ &$v_{1+3}$\\
    \hline
    \multirow{2}{*}{$S_{1+2+3}$} &\multirow{2}{*}{$l_{1+2+3}$} &\multirow{2}{*}{$(k_3^2,\,-\frac{k_3^3}{p_2})$}  &$\frac{k_3p_2}{2}$ &$\widehat{u_{1+2+3}}$\\
      &  &  &$-\frac{k_3^2}{2}$  &$\widehat{v_{1+2+3}}$\\
   \Xhline{1pt}
  \end{tabular}
  \caption*{\captionsetup{justification=raggedright,singlelinecheck=false,format=hang} \quad 
  The solitons $S_j$ $(j=1+2,\,1+3,\,1+2+3)$ are composed by two components $u_{j}$ and $v_{j}$, and their trajectories are listed by \eqref{l02}.}
\end{table}
\subsection{The asymptotic analysis and stem structures with $k_1=-k_2>k_3>0,\,0<p_1=-p_3<p_2$}
Using the same approach as outlined in the previous section, we arrive at the following proposition.
\begin{prop}\label{prop4.3}
The asymptotic forms of the strong 2-resonant 3-soliton with $k_1=-k_2>k_3>0,\,0<p_1=-p_3<p_2$ are as following:

Before collision ($t\to-\infty$):
\begin{flalign}\label{asy07}
\begin{split}
y\to +\infty,\,&S_2:\quad u_2\approx -\frac{k_2p_2}{2}\sech^2(\frac{\xi_2}{2}),\,v_2\approx -\frac{k_2^2}{2}\sech^2(\frac{\xi_2}{2}),\\
&S_3:\quad u_3\approx -\frac{k_3p_3}{2}\sech^2(\frac{\xi_3}{2}),\,v_3\approx -\frac{k_3^2}{2}\sech^2(\frac{\xi_3}{2}),\\
y\to -\infty,\,&S_{1+2}:\, u_{1+2}\approx -\frac{(k_1+k_2)(p_1+p_2)}{2}\sech^2(\frac{\xi_1+\xi_2}{2}),\,v_{1+2}\approx -\frac{(k_1+k_2)^2}{2}\sech^2(\frac{\xi_1+\xi_2}{2}),\\
x\to -\infty,\,&S_{1+3}:\, u_{1+3}\approx -\frac{(k_1+k_3)(p_1+p_3)}{2}\sech^2(\frac{\xi_1+\xi_3}{2}),\,v_{1+3}\approx -\frac{(k_1+k_3)^2}{2}\sech^2(\frac{\xi_1+\xi_3}{2}).
\end{split}
\end{flalign}

After collision ($t\to+\infty$):
\begin{flalign}\label{asy08}
\begin{split}
y\to +\infty,\,&S_2:\quad \widehat{u_2}\approx -\frac{k_2p_2}{2}\sech^2(\frac{\xi_2+\ln a_{23}}{2}),\,\widehat{v_2}\approx -\frac{k_2^2}{2}\sech^2(\frac{\xi_2}{2}),\\
&S_3:\quad \widehat{u_3}\approx -\frac{k_3p_3}{2}\sech^2(\frac{\xi_3+\ln a_{23}}{2}),\,\widehat{v_3}\approx -\frac{k_3^2}{2}\sech^2(\frac{\xi_3+ \ln a_{23}}{2}),\\
y\to -\infty,\,&S_{1+2}:\,u_{1+2}\approx -\frac{(k_1+k_2)(p_1+p_2)}{2}\sech^2(\frac{\xi_1+\xi_2}{2}),\,v_{1+2}\approx -\frac{(k_1+k_2)^2}{2}\sech^2(\frac{\xi_1+\xi_2}{2}),\\
x\to -\infty,\,&S_{1+3}:\,u_{1+3}\approx -\frac{(k_1+k_3)(p_1+p_3)}{2}\sech^2(\frac{\xi_1+\xi_3}{2}),\,v_{1+3}\approx -\frac{(k_1+k_3)^2}{2}\sech^2(\frac{\xi_1+\xi_3}{2}).
\end{split}
\end{flalign}
\end{prop}

\begin{prop}\label{prop4.4}
The stem structures corresponding to asymptotic forms \eqref{asy07} and \eqref{asy08} are as following:
\begin{flalign}
\begin{split}
t\to-\infty,\,S_1:\quad\,\, &u_1\approx -\frac{k_1p_1}{2}\sech^2(\frac{\xi_1}{2}),\,v_1\approx -\frac{k_1^2}{2}\sech^2(\frac{\xi_1}{2}),\\
t\to+\infty,\,S_{1+2+3}:\,&\widehat{u_{1+2+3}}\approx -\frac{(k_1+k_2+k_3)(p_1+p_2+p_3)}{2}\sech^2(\frac{\xi_1+\xi_2\xi_3+\ln a_{23}}{2}),\\
&\widehat{v_{1+2+3}}\approx -\frac{(k_1+k_2+k_3)^2}{2}\sech^2(\frac{\xi_1+\xi_2+\xi_3}{2}).
\end{split}\label{stem04}
\end{flalign}
\end{prop}

\begin{figure}[h!tb]
	\centering
    \subfigure[$u:\,t=-8$]{\includegraphics[height=3.4cm,width=3.4cm]{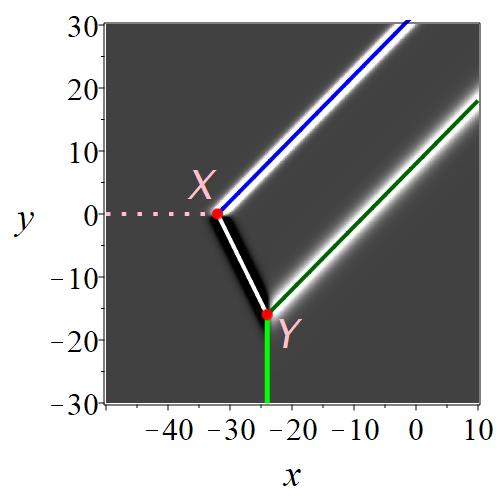}}
    \subfigure[$u:\,t=-4$]{\includegraphics[height=3.4cm,width=3.4cm]{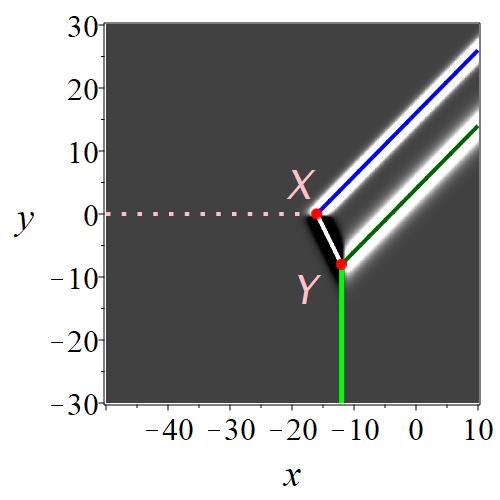}}
	\subfigure[$u:\,t=0$]{\includegraphics[height=3.4cm,width=3.4cm]{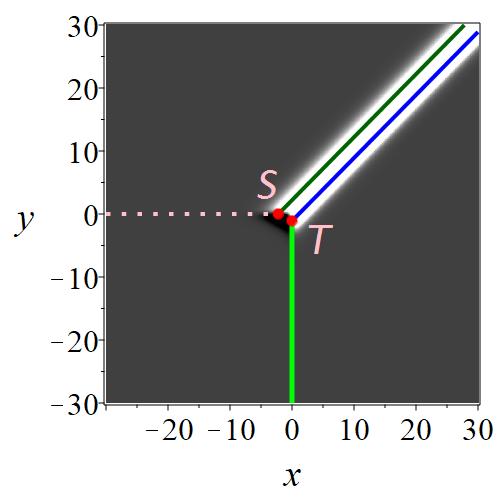}}
	\subfigure[$u:\,t=5$]{\includegraphics[height=3.4cm,width=3.4cm]{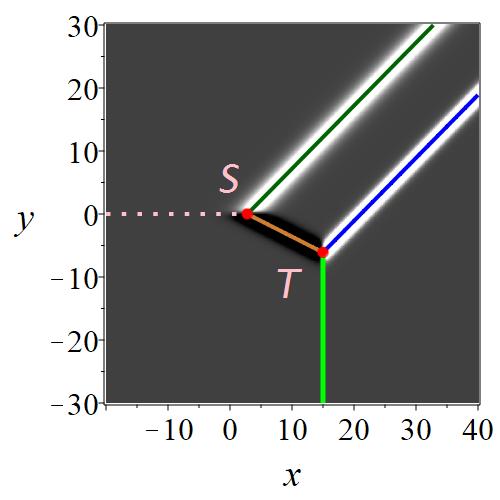}}
    \subfigure[$u:\,t=10$]{\includegraphics[height=3.4cm,width=3.4cm]{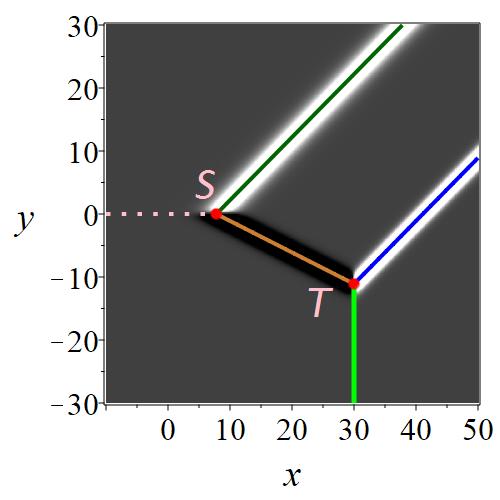}}\\
    \subfigure[$v:\,t=-8$]{\includegraphics[height=3.4cm,width=3.4cm]{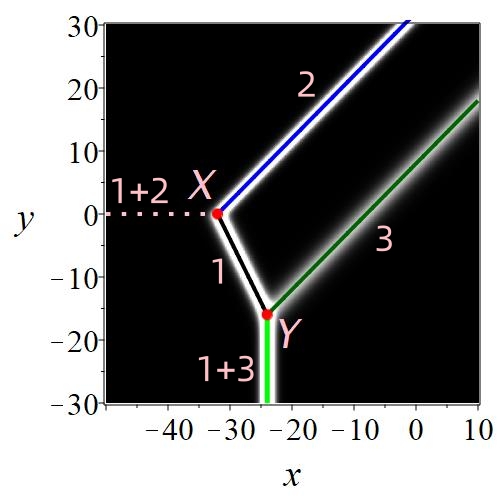}}
    \subfigure[$v:\,t=-4$]{\includegraphics[height=3.4cm,width=3.4cm]{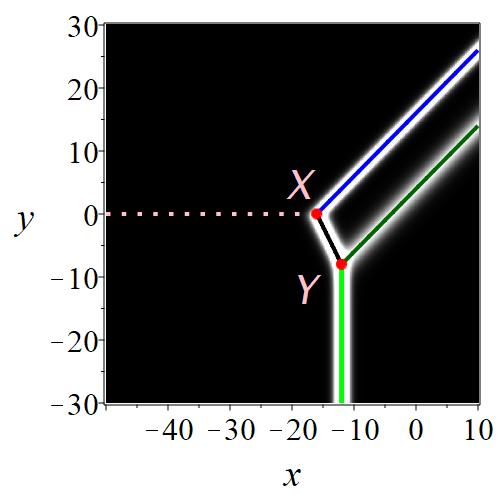}}
	\subfigure[$v:\,t=0$]{\includegraphics[height=3.4cm,width=3.4cm]{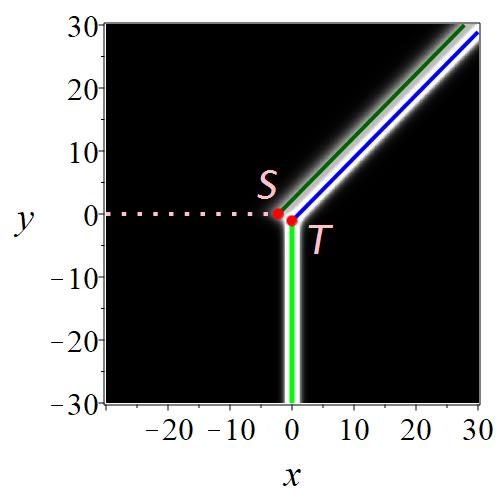}}
	\subfigure[$v:\,t=5$]{\includegraphics[height=3.4cm,width=3.4cm]{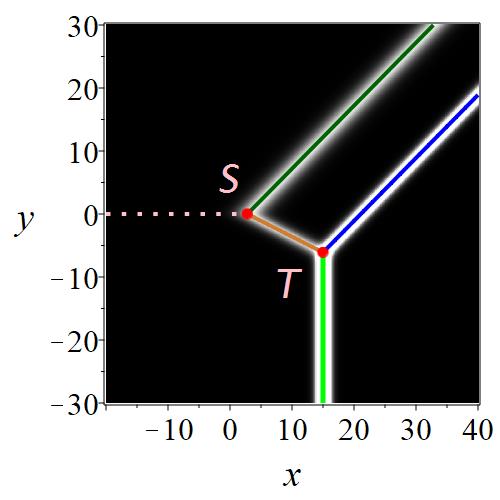}}
    \subfigure[$v:\,t=10$]{\includegraphics[height=3.4cm,width=3.4cm]{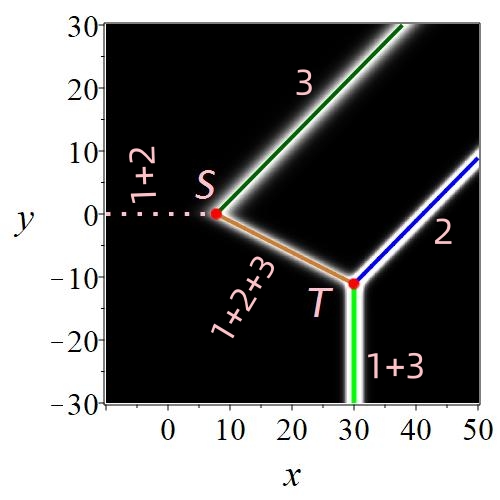}}
	\caption{The density plots of the strong 2-resonant 3-soliton given by \eqref{uv} and \eqref{3soliton2} with $k_1=2,\,k_2=-2,\,k_3=1,\,p_1=1,\,p_2=2,\,p_3=-1,\,\xi_1^0=0,\,\xi_2^0=0,\,\xi_3^0=0$. 
  The lines are the trajectories of the arms and stem structures, and the points are the endpoints of the variable length stem structures.
	}\label{fig3-6}
\end{figure}

The trajectory, amplitude, and velocity of the arms are presented in detail in Tables \ref{tab:t1} and \ref{tab:t5}. 
It is evident from the data that the amplitudes of $v_{1+2}$, $u_{1+2}$, and $u_{1+3}$ are all zero, signifying that $u$ displays three arms while $v$ shows four arms. 
The background plane is represented by density plots, providing a visual context, with the trajectories of the arms indicated by lines. 
Initially, as $t \to -\infty$, $S_1$ appears as a stem structure of variable length. 
As time progresses, the length of $S_1$ decreases, and the distances between $S_2$, $S_3$, and $S_{1+3}$ reduce. 
Near $t=0$, $S_1$ disappears, giving rise to a new stem structure $S_{1+2+3}$, while $S_2$ and $S_3$ continue their movement. 
During this transformation, the arms of $v$ undergo exchanges and reconnections. 
Early on, as $t \to +\infty$, one end of the variable length stem structure $S_1$ connects with $S_{1+2}$, while the other end links with $S_3$ and $S_{1+3}$. 
As time advances toward $t \gg 0$, the stem structure $S_{1+2+3}$ connects with $S_3$ at one end, and with $S_2$ and $S_{1+3}$ at the other. 
The endpoints of these variable length stem structures can be determined using the same method as previously described, resulting in expressions identical to \eqref{STXY}, illustrated as red points in Fig.\ \ref{fig3-6}. 
Consequently, the lengths of the trajectories of these stem structures remain consistent with Eq.\ \eqref{ST+XY}.

Next, we explore the temporal evolution of the amplitudes of the variable length stem structures. 
Similarly to the preceding section, next we will focus on discussing $v$ for simplicity and the formula corresponding to $u$ are provided in the appendix. 
To facilitate computation, we set $k_1=2$, $k_2=-2$, $k_3=1$, $p_1=1$, $p_2=2$, $p_3=-1$, $\xi_1^0=0$, $\xi_2^0=0$, and $\xi_3^0=0$.  
The cross-sectional curves of 3-soliton \eqref{uv} with Eqs.\ \eqref{3soliton2} along $l_1$ and $\widehat{l_{1+2+3}}$ shown in Fig. \ref{fig3-7} (c) and (d) are expressed as,
\begin{flalign}\label{cross34}
\begin{split}
&v|_{l_1}^{(2)}=-18\,{\frac {59049\,{\e^{12\,t+6\,y}}+59049\,{\e^{12\,t+9\,y}}+1053\,{\e^{6\,t}}+4374\,{\e^{6\,t+3\,y}}+2916\,{\e^{
6\,t+6\,y}}+{\e^{-3\,y}}}{ \left( 9+1458\,{\e^{6\,t+3\,y}}+729\,{\e^{6\,t+6\,y}}+{\e^{-3\,y}} \right)^2}},\\
&v|_{\widehat{l_{1+2+3}}}^{(2)}=-{\frac {54\,{\e^{15\,t-3\,x}}+18\,{\e^{39\,t-9\,x}}+72\,{\e^{6\,t}}+8\,{\e^{24\,t-6\,x}}+20\,{\e^{-9\,t+3\,x}}+8
}{ \left( 2+{\e^{24\,t-6\,x}}+{\e^{-9\,t+3\,x}}+9\,{\e^{15\,t-3\,x}} \right) ^2}}.
\end{split}
\end{flalign}

It is worth noting that the amplitudes under investigation in this context are the function values at points $R_5$ and $R_6$. 
However, to avoid confusion with the previous discussion, we opt to re-designate these two points as $R_7$ and $R_{8}$. 
Consequently, the amplitudes of $R_7$ and $R_{8}$ are given by:
\begin{flalign}\label{amplocal4}
\begin{split}
S_{1+2+3}:\quad &v(R_7)=-{\frac{6{\rm e}^{\frac{3t}{2}} \left( 81\,\sqrt {3}{{\rm e}^{6t}}+729\,{{\rm e}^{\frac{15t}{2}}}+162\,\sqrt {3}{{\rm e}^{3t}}+351\,{{\rm e}^
{\frac{9t}{2}}}+\sqrt{3}+36{{\rm e}^{\frac{3t}{2}}}\right) }{\left( 54\,\sqrt {3}{{\rm e}^{\frac{9t}{2}}}+\sqrt {3}{{\rm e}^{\frac{3t}{2}}}+9\,{{\rm e}^{3t}}+3 \right) ^{2}}},\\
S_1:\quad &v(R_8)=-\frac{18{{\rm e}^{\frac{15t}{2}}}+72\,{{\rm e}^{6\,t}}+54\,{{\rm e}^{\frac{9t}{2}}}+8\,{{\rm e}^{3t}}+20\,{{\rm e}^{\frac{3t}{2}}}+8}{\left( 2+{{\rm e}^{3t}}+{{\rm e}^{{\frac{3t}{2}}}}+9\,{{\rm e}^{\frac{9t}{2}}}\right) ^2}.
\end{split}
\end{flalign}
Easily obtain the limits as
$$\lim_{t\to +\infty} v(R_7)=-\frac{1}{2},\,\lim_{t\to -\infty} v(R_7)=0,\,\lim_{t\to +\infty} v(R_8)=0,\,\lim_{t\to -\infty} v(R_8)=-2.$$
This can also be confirmed in Fig.\ \ref{fig3-8}. The plots of $v(R_7)$ and $v(R_8)$ are depicted in Fig.\ \ref{fig3-8} (d). 
It is evident that the evolution of variable length stem structures around $t=0$ is intricate. 
The figure also corroborates the disappearance of $S_1$ around $t=0$ (indicated by $v(R_8)\approx 0$), coinciding with the emergence of $S_{1+2+3}$ ($v(R_7)\approx 0$).

\section{Conclusions and discussions}\label{summary}
This paper systematically studies the asymptotic forms and variable-length stem structures during soliton reconnection, specifically occurring in 2-resonance 3-solitons of the ANNV system \eqref{annveq}. 
The construction method of the 2-resonant 3-soliton solution in this paper is different from that in Ref.\ \cite{interaction12}, but similar to the solution in Ref.\ \cite{interaction12} (see its section 7.2: partial resonance solution). 
During soliton reconnection, two pairs of V-shaped solitons gradually approach, bounce off in another direction, and exhibit the gradual disappearance of one variable-length stem structure alongside the emergence of another (see Figs.\ \ref{fig3-1}, \ref{fig3-2}, \ref{fig3-5}, \ref{fig3-6}). 
We address two distinct cases of 2-resonances: $a_{12}=a_{13}=0,\,0<a_{23}<+\infty$ (weak 2-resonance) and $a_{12}=a_{13}=+\infty,\,0<a_{23}<+\infty$ (strong 2-resonance).

Using two-variable asymptotic analysis method (see paragraph 3, section 3.1), we analytically derive the asymptotic forms of the four arms and stem structures of the 2-resonant 3-solitons as $t \to \pm \infty$. 
The weak 2-resonance case is presented in Propositions \ref{prop3.1}--\ref{prop3.4}, while the strong 2-resonance case is introduced in Propositions \ref{prop4.1}--\ref{prop4.4}. 
This analysis also provides a systematic mathematical theory of soliton reconnection. 
These analytical forms include formulas of the trajectories, amplitudes, and velocities of the arms and stem structures. 
By combining the corresponding trajectories, we determine the vertices of the V-shaped solitons and define them as the endpoints of the stem structures, as shown in Eqs.\ \eqref{EFGH}, \eqref{PQMN}, and \eqref{STXY}. 
We then provide formulas for the length of the stem structures in Eqs.\ \eqref{EF+GH}, \eqref{PQ+MN}, and \eqref{ST+XY}, reflecting the linear evolution of the stem lengths over time for $t \gg 0$ and $t \ll 0$.

Based on the asymptotic forms and stem structures studied in this paper, we observe the following facts:
\begin{itemize}
  \item The 2-resonant 3-solitons of Eq. \eqref{annveq} possess four infinitely extending arms. 
  Especially, owing to the constraint $0 < a_{23} < +\infty$, the arms $S_2$ and $S_3$ manifest a finite phase shift as $t\to\pm\infty$. 
  Conversely, the remaining two arms remain unchanged throughout the collision ($t\to\pm\infty$). 
  This represents a significant improvement over previous studies that only considered spatial progressive behavior.
  \item The 2-resonant 3-solitons of Eq. \eqref{annveq} exhibit distinct stem structures as $t\to-\infty$ and $+\infty$ respectively. 
  Unlike the characteristics observed in the stem structures of quasi-resonant 2-solitons examined in Ref. \cite{yuan2024}, the local properties of the stem structure in soliton reconnections undergo temporal variations.
  \item The profile curve where the trajectory of the conventional stem structure (with non-zero amplitude) is located contains only extreme points, unlike the extreme lines found in line solitons. 
  The location of these extreme points changes over time.
\end{itemize}

A natural extension of the present work would be to investigate partial and complete resonances of higher-order solitons. 
Increasing the number of $a_{ij}$ parameters would give rise to a richer array of partial resonance cases, promising the emergence of more intriguing phenomena. We intend to explore this extension in future studies to provide a more in-depth understanding of the complexities inherent in the ANNV system and to extend our findings to other soliton equations.

\vspace{\baselineskip} 
{{\bf Conflict statement}
	{The authors declare that they have no conflict of interests.}}

\vspace{\baselineskip}
{{\bf Data availability}
{The data that support the findings of this study are available within the article.}}

\vspace{\baselineskip} 
{{\bf Acknowledgments}
	{This work is supported by the National Natural Science Foundation of China (Grants 12071304, 12471239), 
  NUPTSF (Grant NY222169), Shenzhen Natural Science Fund (the Stable Support Plan Program) (Grant 20220809163103001), 
  and Guangdong Basic and Applied Basic Research Foundation (Grant 2024A1515013106)}}
\section*{Appendix}\addcontentsline{toc}{section}{Appendix}
The cross-sectional curves of 3-soliton \eqref{uv} with Eqs.\ \eqref{3soliton1} along $l_{1-2}$ and $l_{1-3}$ shown in Fig.\ \ref{fig3-3} (a) and (b) are expressed as,
\begin{flalign}\label{crossadd1}
\begin{split}
&u|_{l_{1-2}}^{(1)}=-\frac{13}{6}\cdot\frac{60\e^{\frac{15t}{2}+\frac{11y}{3}}+108\e^{\frac{15t}{2}+\frac{17y}{6}}
+390\e^{5t+\frac{13y}{6}}+585\e^{\frac{5t}{2}+\frac{3y}{2}}+52\e^{\frac{5t}{2}+\frac{2y}{3}}}
{(3\e^{5t+\frac{13y}{6}}+26\e^{\frac{5t}{2}+\frac{3y}{2}}+13\e^{\frac{5t}{2}+\frac{2y}{3}}+13)^2},\\
&u|_{l_{1-3}}^{(1)}=-\frac{13}{6}\cdot\frac{156\e^{3x-\frac{33t}{4}}+12\e^{\frac{9x}{2}-\frac{129t}{8}}+390\e^{\frac{5x}{2}-\frac{65t}{8}}+169\e^{\frac{x}{2}-\frac{t}{8}}+468\e^{2x-8t}}
{(3\e^{\frac{5x}{2}-\frac{65t}{8}}+26\e^{\frac{x}{2}-\frac{t}{8}}+13\e^{2x-8t}+13)^2}.
\end{split}
\end{flalign}

The amplitudes of $S_{1-2}$ and $S_{1-3}$ are expressed as:
\begin{flalign}\label{amplocaladd1}
\begin{split}
S_{1-2}:\, &u(R_1)=-\frac{3}{2}\cdot\frac{468\gamma_2\e^{\frac{325t}{144}} +585\gamma_2\e^{\frac{125t}{144}} + 12 \gamma_4\e^{\frac{25t}{8}} + 
    390\gamma_1\e^{\frac{25t}{16}} + 260 \gamma_3}
{\left(3 \gamma_4\e^{\frac{325t}{144}} +3\gamma_1\e^{\frac{25t}{36}} + 9 \e^{\frac{25t}{16}} + 26\gamma_2\right)^2},\\
S_{1-3}:\, &u(R_2)=-\frac{2197\gamma_4\e^{\frac{75t}{16}} + 2028\gamma_3\e^{\frac{45t}{8}} + 5070\gamma_1\e^{\frac{75t}{16}} + 6084\gamma_4\e^{\frac{60t}{16}} + 676\gamma_2\e^{\frac{15t}{16}}}
    {2\left(26\gamma_2\e^{\frac{75t}{16}} +39 \e^{\frac{15t}{4}} +3\gamma_1\e^{\frac{15t}{16}} + 13\gamma_4 \right)^2},
\end{split}
\end{flalign}
where $\gamma_1=3^{\frac{3}{8}}\cdot 13^{\frac{5}{8}},\,\gamma_2=3^{\frac{7}{8}}\cdot 13^{\frac{1}{8}},\,\gamma_3=3^{\frac{1}{4}}\cdot 13^{\frac{3}{4}},\,\gamma_4=\sqrt{39}$.

The cross-sectional curves of 3-soliton \eqref{uv} with Eqs.\ \eqref{3soliton1} along $l_{1-2}$ and $l_{1-3}$ are expressed as,
\begin{flalign}\label{crossadd2}
\begin{split}
&u|_{l_{1-2}}^{(2)}=-\frac{75 \e^{5x - 17t} + 90 \e^{4x - 10t} + 450 \e^{3x - 9t} + 2250 \e^{2x - 8t} + 450 \e^{x - t}}
{\left( \e^{3x - 9t} + 30 \e^{2x - 8t} + 15 \e^{x - t} + 15 \right)^2},\\
&u|_{l_{1-3}}^{(2)}=-\frac{75 \e^{5x - 17t} + 90 \e^{4x - 10t} + 450 \e^{3x - 9t} + 2250 \e^{2x - 8t} + 450 \e^{x - t}}
{\left( \e^{3x - 9t} + 30 \e^{2x - 8t} + 15 \e^{x - t} + 15 \right)^2}.
\end{split}
\end{flalign}

The amplitudes of the variable length stem structures, respectively, are
\begin{flalign}\label{amplocaladd2}
\begin{split}
S_{1-2}:\quad &u(R_3)=-\frac{90 \e^{6t} + 2 \alpha_6\e^{\frac{15t}{2}} + 30\alpha_6\e^{\frac{9t}{2}} +150 \e^{3t} + 75\alpha_6\e^{\frac{3t}{2}}}
{\left(\alpha_6\e^{\frac{9t}{2}} + \alpha_6\e^{\frac{3t}{2}} + \e^{3t} + 30 
    \right)^2},\\
S_{1-3}:\quad &u(R_4)=-\frac{225 \left( 9\alpha_8\e^{\frac{7t}{2}} + 6 \alpha_7\e^{4t} + 2 \alpha_6\alpha_7\e^{\frac{5t}{2}} + 6\alpha_6\e^{\frac{3t}{2}} + 2\alpha_7\e^t \right)}
{\left( 30\alpha_7\e^{\frac{5t}{2}} + \alpha_6\alpha_7\e^t + 15 \e^{\frac{3t}{2}} + 15\alpha_6\right)^2},
\end{split}
\end{flalign}
where $\alpha_6=\sqrt{15},\,\alpha_7=\sqrt[3]{15},\,\alpha_8=\sqrt[6]{15}$.

The cross-sectional curves of 3-soliton \eqref{uv} with Eqs.\ \eqref{3soliton2} along $\widehat{l_{1+2+3}}$ and $l_{1}$ are expressed as,
\begin{flalign}\label{crossadd3}
\begin{split}
&u|_{\widehat{l_{1+2+3}}}^{(1)}=\frac{169 \left( 108 \sqrt{39} \e^{\frac{15t}{4} + \frac{y}{2}} + 52 \beta_1\e^{\frac{15t}{8} + \frac{7y}{2}} + 
273\beta_1\e^{\frac{15t}{8} + \frac{4y}{3}} + 12\beta_1\e^{\frac{15t}{8} - \frac{5y}{6}} +  6084 \e^{\frac{13y}{6}} - 2704 \right)}
{6 \left( 13\beta_1\e^{\frac{15t}{8} + \frac{4y}{3}} + 3 \beta_1\e^{\frac{15t}{8} - \frac{5y}{6}} + 169 \e^{\frac{13y}{6}} + 338 \right)^2},\\
&u|_{l_{1}}^{(1)}=\frac{108 \e^{-\frac{65t}{8} + \frac{5x}{2}} +  468 \e^{-\frac{559t}{72} + \frac{19x}{18}} + 182 \e^{-\frac{143t}{18} + \frac{16x}{9}} + 52 \e^{-\frac{1157t}{72} + \frac{77x}{18}} + 
    12 \e^{\frac{13t}{72} - \frac{13x}{18}} - 27}
    {2 \left( 3 \e^{-\frac{65t}{8} + \frac{5x}{2}} + 13 \e^{-\frac{143t}{18} + \frac{16x}{9}} + 3 \e^{\frac{13t}{72} - \frac{13x}{18}} + 6 \right)^2},
\end{split}
\end{flalign}
where, $\beta_1=13^{\frac{3}{4}} \cdot 3^{\frac{1}{4}}$.

The amplitudes of $R_5$ and $R_6$ for $u$ are given as following,
\begin{flalign}\label{amplocaladd3}
\begin{split}
S_{1+2+3}:\, &u(R_5)=\frac{13 \left(36\beta_2\e^{\frac{93t}{16}} +1404\beta_2\e^{\frac{69t}{16}} + 819\beta_1\e^{\frac{27t}{8}} +1404\beta_3\e^{\frac{39t}{16}} + 676\beta_3\e^{\frac{15t}{16}} - 35152 \right)}
{6 \left(3\beta_1\e^{\frac{27t}{8}} + 3\beta_3\e^{\frac{39t}{16}} + 13\beta_3\e^{\frac{15t}{16}} + 338 \right)^2},\\
S_1:\quad &u(R_6)=\frac{108 \e^{-\frac{15t}{4}} + 468 \e^{-\frac{71t}{12}} + 182 \e^{-\frac{29t}{6}} + 52 \e^{-\frac{103t}{12}} + 12 \e^{-\frac{13t}{12}} - 27}
{2 \left( 3 \e^{-\frac{15t}{4}} + 13 \e^{-\frac{29t}{6}} + 3 \e^{-\frac{13t}{12}} + 6 \right)^2},
\end{split}
\end{flalign}
where, $\beta_1=13^{\frac{3}{4}} \cdot 3^{\frac{1}{4}},\,\beta_2=13^{\frac{1}{8}} \cdot 3^{\frac{7}{8}},\,\beta_3=13^{\frac{3}{8}} \cdot 3^{\frac{5}{8}}$.

The cross-sectional curves of 3-soliton \eqref{uv} with Eqs.\ \eqref{3soliton2} along $l_1$ and $\widehat{l_{1+2+3}}$ for $u$ are expressed as,
\begin{flalign}\label{crossadd4}
\begin{split}
&u|_{l_1}^{(2)}=\frac{
    59049 \e^{3y} + 
    \e^{-12t - 9y} + 
    324 \e^{-6t - 6y} + 
    2916 \e^{-6t} - 118098
}{
    \left( 
        \e^{-6t - 6y} + 
        9 \e^{-6t - 3y} + 
        729 \e^{3y} + 1458 
    \right)^2
},\\
&u|_{\widehat{l_{1+2+3}}}^{(2)}=\frac{
    59049 \e^{3y} + 
    \e^{-12t - 9y} + 
    324 \e^{-6t - 6y} + 
    2916 \e^{-6t} - 118098
}{
    \left( 
        \e^{-6t - 6y} + 
        9 \e^{-6t - 3y} + 
        729 \e^{3y} + 1458 
    \right)^2
}.
\end{split}
\end{flalign}

The amplitudes of $R_7$ and $R_{8}$ are given by:
\begin{flalign}\label{amplocaladd4}
\begin{split}
S_{1+2+3}:\quad &u(R_7)=\frac{2 \sqrt{3} \e^{-\frac{15t}{2}} + 72 \e^{-6t} + 216 \e^{-3t} + 162 \sqrt{3} \e^{-\frac{3t}{2}} - 2916}
{\left(\sqrt{3} \e^{-\frac{9t}{2}} +3 \sqrt{3} \e^{-\frac{3t}{2}} +\e^{-3t} + 54 \right)^2},\\
S_1:\quad &u(R_8)=\frac{72 \e^{6t} + 18 \e^{\frac{15t}{2}} + 9 \e^{\frac{9t}{2}} + 8 \e^{3t} + 2 \e^{\frac{3t}{2}} - 4}
{\left( 9 \e^{\frac{9t}{2}} + \e^{3t} + \e^{\frac{3t}{2}} + 2 \right)^2}.
\end{split}
\end{flalign}


\end{document}